\documentclass{article}

\usepackage{arxiv}

\usepackage[utf8]{inputenc} 
\usepackage[T1]{fontenc}    
\usepackage{hyperref}       
\usepackage{url}            
\usepackage{booktabs}       
\usepackage{amsfonts}       
\usepackage{nicefrac}       
\usepackage{microtype}      
\usepackage{graphicx}
\usepackage{lipsum}
\usepackage{enumitem}
\usepackage[dvipsnames]{xcolor}
\usepackage{multirow}
\usepackage{cancel}
\usepackage{soul,color}

\usepackage{lipsum}
\usepackage{soul}
\usepackage{color}
\usepackage{float}
\usepackage{graphicx}
\usepackage{caption}
\usepackage{subcaption}
\usepackage{titlesec}
\usepackage{multirow}
\usepackage{amsmath}
\usepackage{soul,color}
\usepackage[section]{placeins}
\usepackage{array, makecell} %
\usepackage{float}
\usepackage[section]{placeins}
\usepackage{wrapfig}
\usepackage{enumitem}
\usepackage{amsmath} 

\usepackage{times}
\usepackage{soul}
\usepackage{url}

\usepackage{amsmath}
\usepackage{amsthm}
\usepackage{booktabs}
\usepackage{algorithm}
\usepackage{algpseudocode}
\usepackage{amsfonts}
\usepackage{bm}
\usepackage{amssymb}
\usepackage{chemformula}
\usepackage{makecell}
\usepackage{mathtools}
\usepackage[export]{adjustbox}
\usepackage{multirow}
\usepackage{enumitem}
\usepackage{tabularx}
\usepackage{subcaption}
\usepackage[toc,page]{appendix}

\DeclarePairedDelimiter\floor{\lfloor}{\rfloor}

\newcommand{\norm}[1]{\left\lVert#1\right\rVert}
\newcommand{\etal}{\textit{et al.}}
\usepackage[sort&compress,numbers,square]{natbib}

\theoremstyle{plain}

\theoremstyle{definition}

\usepackage{graphicx, color}
\graphicspath{{fig/}}

\usepackage[numbers,sort&compress]{natbib}

\title{Physics Guided Deep Learning for Generative Design of Crystal Materials with Symmetry Constraints}

\author{
   Yong Zhao \\
  Department of Computer Science and Engineering\\
  University of South Carolina\\
  Columbia, SC, 29201, USA \\
   \And
  Edirisuriya M. Dilanga Siriwardane\\
 Department of Physics\\
  University of Colombo\\
 Colombo 03, Sri Lanka \\
 \And
    Zhenyao Wu, Nihang Fu \\
  Department of Computer Science and Engineering\\
  University of South Carolina\\
  Columbia, SC, 29201, USA \\
 \And
 Mohammed Al-Fahdi, Ming Hu*\\
   Department of Mechanical Engineering\\
  University of South Carolina\\
  Columbia, SC, 29201, USA \\
  \texttt{hu@sc.edu}
  \And
 Jianjun Hu \thanks{Corresponding author: J.H. (http://www.cse.sc.edu/~jianjunh)}\\
  Department of Computer Science and Engineering\\
  University of South Carolina\\
  Columbia, SC, 29201, USA \\
  \texttt{jianjunh@cse.sc.edu} \\
}
\begin{document}
\maketitle

\begin{abstract}
Discovering new materials is a challenging task in materials science crucial to the progress of human society. Conventional approaches based on experiments and simulations are labor-intensive or costly with success heavily depending on experts' heuristic knowledge. Here, we propose a deep learning based Physics Guided Crystal Generative Model (PGCGM) for efficient crystal material design with high structural diversity and symmetry. Our model increases the generation validity by more than 700\% compared to FTCP, one of the latest structure generators and by more than 45\% compared to our previous CubicGAN model. Density Functional Theory (DFT) calculations are used to validate the generated structures with 1,869 materials out of 2,000 are successfully optimized and deposited into the Carolina Materials Database \url{www.carolinamatdb.org}, of which 39.6\% have negative formation energy and 5.3\% have energy-above-hull less than 0.25 eV/atom, indicating their thermodynamic stability and potential synthesizability.

\end{abstract}

\section{Introduction}

Understanding the relationship of structures and functions is one of the most important questions in many disciplines such as chemistry, materials science, biology, which is critical for rational design of structures for achieving specific e.g. molecule, protein, or materials functions. However, the sophistication of the physical, chemical, and geometric atomic interactions makes it challenging to exhaustively enumerate the "design rules" for heuristic design approaches. On the other hand, the lack of sufficient data, the diversity of the hypothetical structure types, along with the strong physicochemical constraints makes it hard even for data driven approaches to structure design.

Here, we propose a physics guided deep learning model for generative design of crystal materials. Solid crystals such as ionic conductors, perovskites, photovoltaics, and piezoelectrics, play an important role in modern industries. Over centuries, humanity has dedicated significant amount of efforts to discovering high-performance functional materials. However, to date, only around 250,000 inorganic materials have been experimentally determined as collected by the ICSD database~\cite{belsky2002new}, which only covers a tiny portion of the almost infinite material design space considering the combinatorial space with the number of elements cross the periodic table and the total 230 possible symmetries of crystal structures. Traditional trial-and-error tinkering methods for materials discovery are mainly reliant on domain experts' knowledge~\cite{noh2019inverse}, which is time-consuming and labor-intensive. To meet the high demand for new functional materials, we need more efficient strategies to explore the vast chemical space to accelerate the materials discovery process.

Currently, {a} popular approach to generating materials is based on element substitution of existing materials combined with high-throughout virtual screening (HTVS)~\cite{pyzer2015high}. The whole process contains three steps: (1) combinatorially substituting elements in known crystal structures, (2) optimizing the candidate structures using density functional theory (DFT) calculations, and (3) experimental verification. Typical large computational materials databases created by HTVS are Materials Project (MP)~\cite{jain2013commentary} and Open Quantum Materials Database (OQMD)~\cite{kirklin2015open}. Despite its promising usage in material design, a fundamental drawback of HTVS is that it cannot generate materials beyond the structural prototypes of existing materials. It is also extremely computationally intensive and its success rate heavily depends on experts' intuitions.

One way to overcome the drawbacks of HTVS for discovering materials is to perform crystal structure prediction for candidate material compositions using global optimization techniques, which are used to identify their stable structural phases. Simulated annealing has been used to predict the structures of alloys~\cite{franceschetti1999inverse} and boron nitride~\cite{doll2008structure}. The minima hopping ~\cite{amsler2010crystal} is another algorithm for finding unknown crystalline structures \cite{flores2020crystal}. {Two} widely used crystal structure prediction (CSP) algorithms are USPEX~\cite{glass2006uspex} and CALYPSO~\cite{wang2012calypso}, which use evolutionary algorithms and particle swarm optimization for finding crystal structures. Despite their success in a variety of cases, these CSP based approaches for  materials discovery suffer from their limited applicability to only relative simple structures usually with small number of atoms in unit cell.

Another promising approach to design solid materials beyond known crystal structure prototypes is generative deep learning models \cite{zhao2021high,fuhr2022deep,schwalbe2020generative}, which can learn data distribution (knowledge of forming stable crystal structures) from known materials and then sample from it to generate materials. Variational Auto-encoder (VAE)~\cite{kingma2013auto} and Generative Adversarial Network (GAN)~\cite{goodfellow2014generative} are two popular generative models used to generate materials. 
A VAE~\cite{kingma2013auto} model consists of two deep neural networks, an encoder and a decoder. The encoder is trained to encode materials into latent vectors and the decoder reconstructs the materials from the latent vectors. After training, different strategies can be used to sample the latent space and then use the decoder to generate materials. iMatGen~\cite{noh2019inverse} is believed to be the first work that uses VAE to realize the inverse design of solid materials. It encodes unit cells into 3D grid based representations, and spherical linear interpolation and Gaussian random sampling are used to sample from the latent space to generate materials. Hoffmann \etal~\cite{hoffmann2019data} extend iMatGen by combining a UNet module to segment reconstructed 3D voxel images into atoms. Based on iMatGen and Hoffmann \etal, ICSG3D~\cite{court20203} integrates formation energy per atom into 3D voxelized solid crystals, which enables the VAE to encode materials and energy simultaneously. This makes it possible to generate materials subject to user-defined formation condition. Another approach to represent 3D crystals is to encode 2D crystallographic representations as the combination of the real space and the reciprocal-space Fourier-transformed features~\cite{ren2020inverse}. In CDVAE~\cite{xie2021crystal}, a diffusion network is trained to generate material structures ~\cite{ho2020denoising}, in which a diffusion process within their diffusion variational autoencoder moves atoms into positions in the lower energy space to generate stable crystals. All these models have difficulty in generation of high quality structures with high symmetry (e.g. space group number >=62) due to their neglecting the structure symmetry in their generation models, a major special characteristic of periodic crystal structures. 
A GAN model~\cite{goodfellow2014generative} consist of two deep neural networks, a generator and a discriminator (critic). The generator creates fake materials with inputs of random vectors with or without conditioning on elements and space groups while the discriminator tries to tell real materials from generated ones. With learnt knowledge of forming crystals, the generator can directly create materials. The first method to generate materials using GAN is CrystalGAN~\cite{nouira2019crystalgan}, which leverages a CycleGAN~\cite{zhu2017unpaired} to generate ternary materials from existing binaries. However, it remains uncertain whether CrystalGAN can be extended to produce more complex crystals. GANCSP~\cite{kim2020generative} and CubicGAN~\cite{zhao2021high} are two GAN based generation models that directly encode crystal structures as matrices containing information of fractional coordinates, element properties, and lattice parameters, which are fed as inputs to build models that generate crystals conditioned on composition or both composition and space group. The major difference between them is that GANCSP can only generate structures of a specific chemical system (e.g. Mg-Mn-O system) while CubicGAN can generate structures of diverse systems of three cubic space groups. In CCDCGAN~\cite{long2021constrained}, Long \etal use 3D voxelized crystals as inputs for their autoencoder model, which then converts them to 2D crystal graphs, which is used as the inputs to the GAN model. A formation energy based constraint module is trained with the discriminator, which automatically guides the search for local minima in the latent space. More recently, modern generative models such as normalizing flow \cite{ahmad2022free,wirnsberger2022normalizing} and diffusion models have also been \cite{xie2021crystal} (CDVAE) or planned to be \cite{baird2022xtal2png} applied to crystal structure generation. Less related works include MatGAN~\cite{dan2020generative} and CondGAN($x^{bp}$)~\cite{sawada2019study} developed for generating only chemical compositions.

Despite the success of VAEs and GANs in material generation \cite{noh2019inverse,zhao2021high,xie2021crystal}, all current generative models have several major drawbacks. For example, the iMatGen algorithm ~\cite{noh2019inverse} can only generate structures of a specific chemical system such as vanadium oxides and only several metastable VxOy materials were discovered out of 20,000 generated hypothetical materials. Similarly, GANCSP ~\cite{kim2020generative} and CrystalGAN~\cite{nouira2019crystalgan} only generate for a given chemical system (e.g. Mg-Mn-O system and hydride systems). VAE-UNet pipeline developed in~\cite{court20203} expands the diversity of generated materials and can reconstruct the atom coordinates more accurately by incorporating UNet segmentation and conditioning on properties. However, VAE-UNet still confines itself to cubic crystal system generation and the number of atoms in a unit cell is restricted to no more than 40. All above discussed works do not realize high-throughout generation of crystal materials. CubicGAN~\cite{zhao2021high} is an early public example of a high-throughput generative deep learning model for (cubic) crystal structures, which has discovered four prototypes with 506 materials confirmed to be stable by DFT calculations. Although CubicGAN has generated millions of crystal structures with hundreds of stable ones confirmed, the generated structures are limited to three space groups in the cubic crystal system, of which the atom coordinates are assumed to be multiples of 1/4: it is not capable of generating generic atom coordinates. While these works open the door to generative design of materials, several unique challenges still remain that prevents effective generative design: (1) how to learn the physical atomic constraints of stable materials to enable efficient sampling; (2) how to achieve precise generation of atom fractional coordinates and lattice parameters; (3) how to handle the extreme bias of the distribution of materials in the 230 space groups; (4) how to exploit the high symmetry of crystal structures in the generation process.

In this work. we introduce a physics guided crystal generative model (PGCGM) to exploit the physical rules for addressing aforementioned challenges. Our contributions are summarized as follows:

\begin{enumerate}
    \item We present a physics guided deep generative model for crystal generation that combines the space group affine transformation and an efficient self-augmentation method.
    \item We propose two physics-oriented losses based on atomic pairwise distance constraints and structural symmetry to fuse the physical laws into deep learning model training.
    \item We evaluate our model against two baselines to show its superiority and perform DFT calculations to validate our generated structures with high success rate (93.5\% can be optimized successfully).
\end{enumerate}

\section{Results and Discussion}

\paragraph{Dataset}{We collect our material data from MP~\cite{jain2013commentary}, ICSD~\cite{belsky2002new} and OQMD (v1.4)~\cite{kirklin2015open}. In total, 42,072 ternary materials with 20 space groups {(5 crystal systems)} are curated when we start this project (The choice of 20 space groups is due to the limited structure samples for other remaining space groups. Our model is applicable to any space group given sufficient training samples). We use a 80-20 random training/validation split for all of our experiments. We term the dataset with 42,072 materials as \textbf{MIO}. When conducting this project, the latest version of OQMD is just yet released. There are 9,441 ternary materials that are filtered by the same criteria and are brand-new materials in the latest OQMD (v1.5). We use these 9,441 ternary materials as our test dataset \textbf{TST} to compare our method with two baselines. Details regarding dataset collection are in Dataset Curation section of supplementary materials.}

\paragraph{Generation performance}
We compare PGCGM with two latest algorithms that can generate crystals with multiple chemical systems instead of only a special group of materials (e.g. VxOy and Mg-Mn-O systems)~\cite{noh2019inverse,kim2020generative}. FTCP~\cite{ren2020inverse} combines real space properties (e.g., atom coordinates) and momentum-space properties (e.g. diffraction pattern) to represent crystal structures. Then a CNN based VAE is trained for materials generation. CubicGAN~\cite{zhao2021high} trains a WGAN-GP~\cite{gulrajani2017improved} to generate cubic structures in three space groups and here we expand the original method to 20 space groups.

\begin{table}[H]
    \centering
    \caption{Material generation performance. \textit{PGCGM+dist} is the model with atom distance based loss. \textit{PGCGM+dist+coor} is the model with atom distance based loss and coordinates based loss.}
    \begin{tabular}{|cc|ccc|ccc|c|}
    \hline
    \multicolumn{2}{|c|}{\multirow{2}{*}{Method}}    & \multicolumn{3}{c|}{Validity (\%)}    & \multicolumn{3}{c|}{Prop. Dist.}                            & (\%) \\ \cline{3-9}
    
    \multicolumn{2}{|c|}{}                     & \multicolumn{1}{c|}{CIFs} &\multicolumn{1}{c|}{Distance} &Charge  & \multicolumn{1}{c|}{minD} & \multicolumn{1}{c|}{maxD} &density  &Diversity  \\ \hline
    
    \multicolumn{2}{|c|}{FTCP}                     & \multicolumn{1}{c|}{0.88} &\multicolumn{1}{c|}{63.28} &49.89  & \multicolumn{1}{c|}{1.685} & \multicolumn{1}{c|}{0.754} &2.895 &89.9  \\ \hline
    
    \multicolumn{2}{|c|}{CubicGAN}                     & \multicolumn{1}{c|}{4.97} &\multicolumn{1}{c|}{99.0} &59.47  & \multicolumn{1}{c|}{0.626} & \multicolumn{1}{c|}{3.476} &3.871  &98.0  \\ \hline
    
    \multicolumn{2}{|c|}{{PGCGM}} & \multicolumn{1}{c|}{1.98} &\multicolumn{1}{c|}{\textbf{99.54}} &57.36  & \multicolumn{1}{c|}{\textbf{0.224}} & \multicolumn{1}{c|}{3.664} &2.675  &\textbf{98.4}  \\ \hline
    
    \multicolumn{2}{|c|}{PGCGM+dist}& \multicolumn{1}{c|}{\textbf{7.14}} &\multicolumn{1}{c|}{99.47}&61.82  & \multicolumn{1}{c|}{0.405} & \multicolumn{1}{c|}{0.520} & \textbf{0.765} &96.3  \\ \hline
    
    \multicolumn{2}{|c|}{PGCGM+dist+coor} & \multicolumn{1}{c|}{6.07} &\multicolumn{1}{c|}{99.43}&\textbf{63.34}  & \multicolumn{1}{c|}{0.357} & \multicolumn{1}{c|}{\textbf{0.490}} &0.791  &97.0  \\ \hline
    
    \end{tabular}
    \label{tab:res}
\end{table}

The performance is shown in Table~\ref{tab:res}. For each method, we sample 500,000 structures and for PGCGM and CubicGAN, we perform atom clustering and merging. However, our atom clustering and merging cannot proceed with materials generated by FTCP and then we did not perform atom clustering and merging on those materials. The percentage of Crystallographic Information Files (CIFs) that are readable by \texttt{pymatgen}~\cite{ong2013python} are shown in the \textit{CIFs} column. Here readable means it can be proceeded by \texttt{pymatgen.core.structure.Structure.from\_file}.  We can find that PGCGM+dist has the largest percentage of materials left and PGCGM+dist+coor comes next. It tells us that distance and coordinates losses play a big part in generating readable materials. For later percentage related metrics, we use the number of CIFs left of each method as denominator. Our model significantly outperforms FTCP by 36.4\% in terms of distance validity and is slightly better than CubicGAN. In terms of distance validty, our model outperforms FTCP and CubicGAN by 6.5\% and 27.0\%, respectively. Since validity are relatively weak metrics, property distribution is further used to provide a stronger metric to evaluate whether the generated materials are realistic. Our model significantly outperforms both two baselines. In terms of minimum atom distance, PGCGM decreases wasserstein distance (WD) by 1.461 compared to FTCP and by 0.402 compared to CubicGAN. In terms of maximum atom distance, PGCGM+dist+coor decreases WD by 0.264 compared to FTCP and by 2.986 compared to CubicGAN. Although CubicGAN has a close minimum atom distance distribution to PGCGM, the much bigger gap of maximum atom distance distribution between CubicGAN and PGCGM+dist+coor indicates that CubicGAN tends to generate larger crystal structures. In terms of density, PGCGM+dist decreases WD by 2.130 compared to FTCP and by 3.106 compared to CubicGAN. PGCGM also achieves the best diversity score even though it generates more readable CIFs than FTCP, which further shows that FTCP is not able to generate not only physically realistic materials but also materials with restricted diversity of formulas. We choose PGCGM+dist+coor as our finalized model to generate materials for late analysis since PGCGM+dist+coor has better properties distribution performance than PGCGM and PGCGM+dist on average. 

\paragraph{Analysis of materials optimized by Bayesian Optimization with Symmetry Relaxation}
Bayesian optimization with Symmetry Relaxation (BOWSR) algorithm~\cite{zuo2021accelerating} is an approach that uses Bayesian optimization to iteratively search lower energy surface to optimize the crystal structures based on the properties predicted by deep learning methods, such as CGCNN~\cite{xie2018crystal} and MEGNet~\cite{chen2019graph}. Instead of directly using expensive DFT for relaxing generated materials, we first use BOWSR to optimize structures generated by our model and two baseline models and then use DFT calculation to further relax them. We randomly select 2,000 generated materials with less than or equal to 32 atoms for FTCP, CubicGAN and PGCGM. We select 100 materials for 20 space groups equally generated by PGCGM. Note that we also use the same 2,000 materials of PGCGM for further DFT analysis. Because some space groups are underrepresented (with less than 100 materials) in CubicGAN-generated materials, we select all materials under these space groups and then we select materials for the rest of space groups proportionally to obtain 2000 materials. For FTCP, materials that can be successfully analyzed to have space groups by \texttt{pymatgen get\_space\_group\_info} with \texttt{symprec=0.1}~\cite{ong2013python} surprisingly all belong to space group P1, which means FTCP loses the significant symmetric constraints when generating materials. Our methods PGCGM and CubicGAN are much better than FTCP in terms of space groups retention. Moreover, it takes more than 10 times time to optimize materials generated by FTCP than by PGCGM and CubicGAN using BOWSR. We use \texttt{StructureMatcher} from \texttt{pymatgen}~\cite{ong2013python} to match the generated materials with the corresponding optimized materials by BOWSR.

\begin{table}[H]\
\centering
\caption{Bayesian optimization performance}
\begin{tabular}{|c|c|c|c|}
\hline
 Method& \# of optimized materials & Match rate (\%) & RMS \\ \hline
 FTCP& 1982 & 0.8 & \textbf{0.006} \\ \hline
 CubicGAN& 1992 & 26.0 & 0.035 \\ \hline
 PGCGM(ours)& \textbf{1994} & \textbf{34.9} & 0.052 \\ \hline
\end{tabular}
\label{tab:bomat}
\end{table}

Table~\ref{tab:bomat} shows the match rate and RMS displacement. The match rate is the percentage of materials satisfying the criteria \texttt{ltol=0.2, stol=0.2, angle\_tol=0.5} and then we calculate RMS displacement for the matched materials. Firstly we find that our method has a slightly higher number of successfully optimized materials by BOWSR. However, our method significantly outperforms FTCP and CubicGAN by 4,200\% and 34.23\% in terms of match rate, respectively. It seems that FTCP has the best RMS displace but the extremely low match rate might tell us that BOWSR is hard to optimize materials with low symmetry, such as space group P1. CubicGAN comes next in terms of RMS displacement.

\begin{figure}[H]
    \centering
    \captionsetup{justification=centering}
    \begin{subfigure}{0.4\textwidth}
        \centering
        \includegraphics[width=0.85\linewidth]{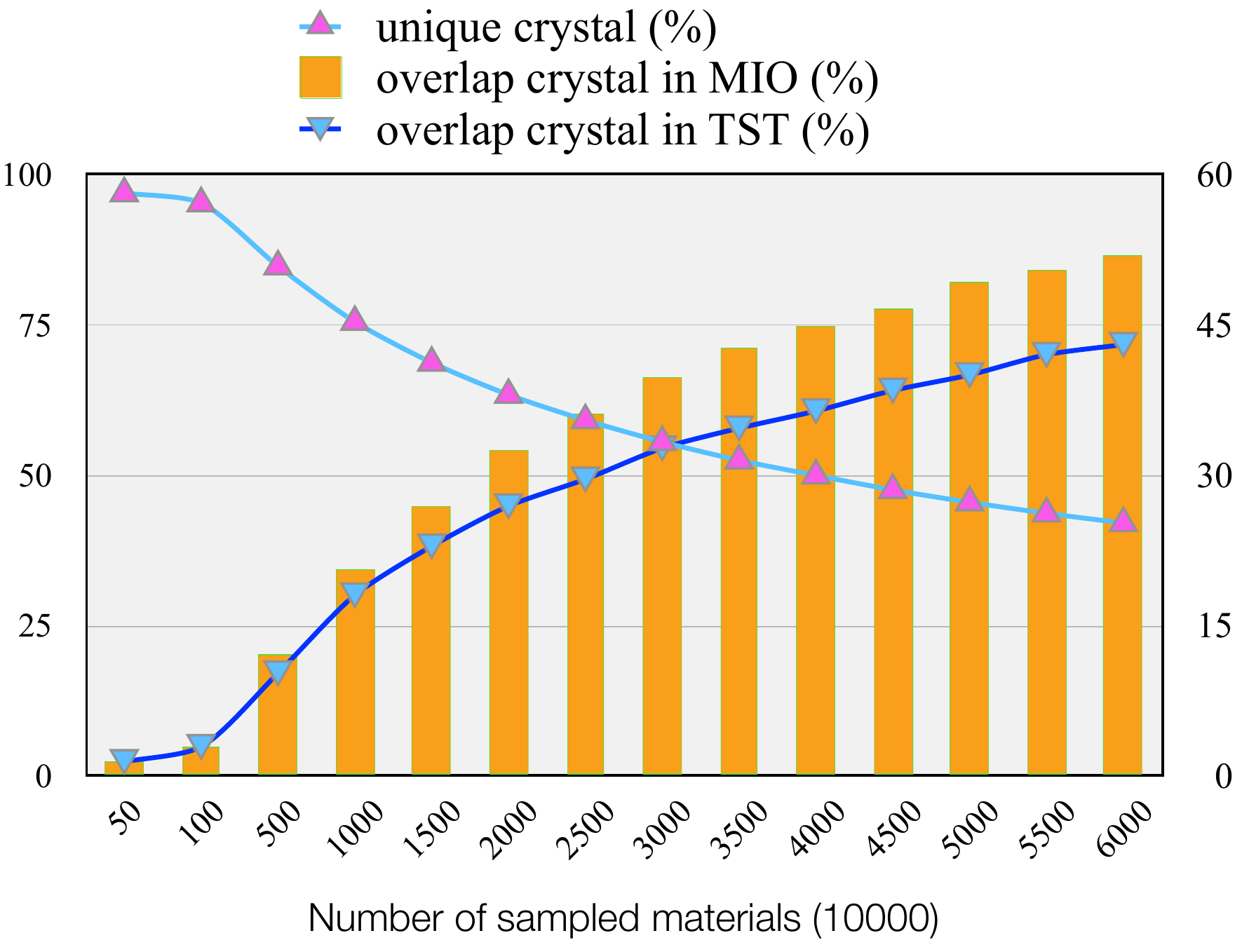}	
      	\caption{The rediscover rate in training and test datasets.}
      	\label{fig:rdcrate}
    \end{subfigure}\hfill
    \begin{subfigure}{0.6\textwidth}
        \centering
        \includegraphics[width=0.85\linewidth]{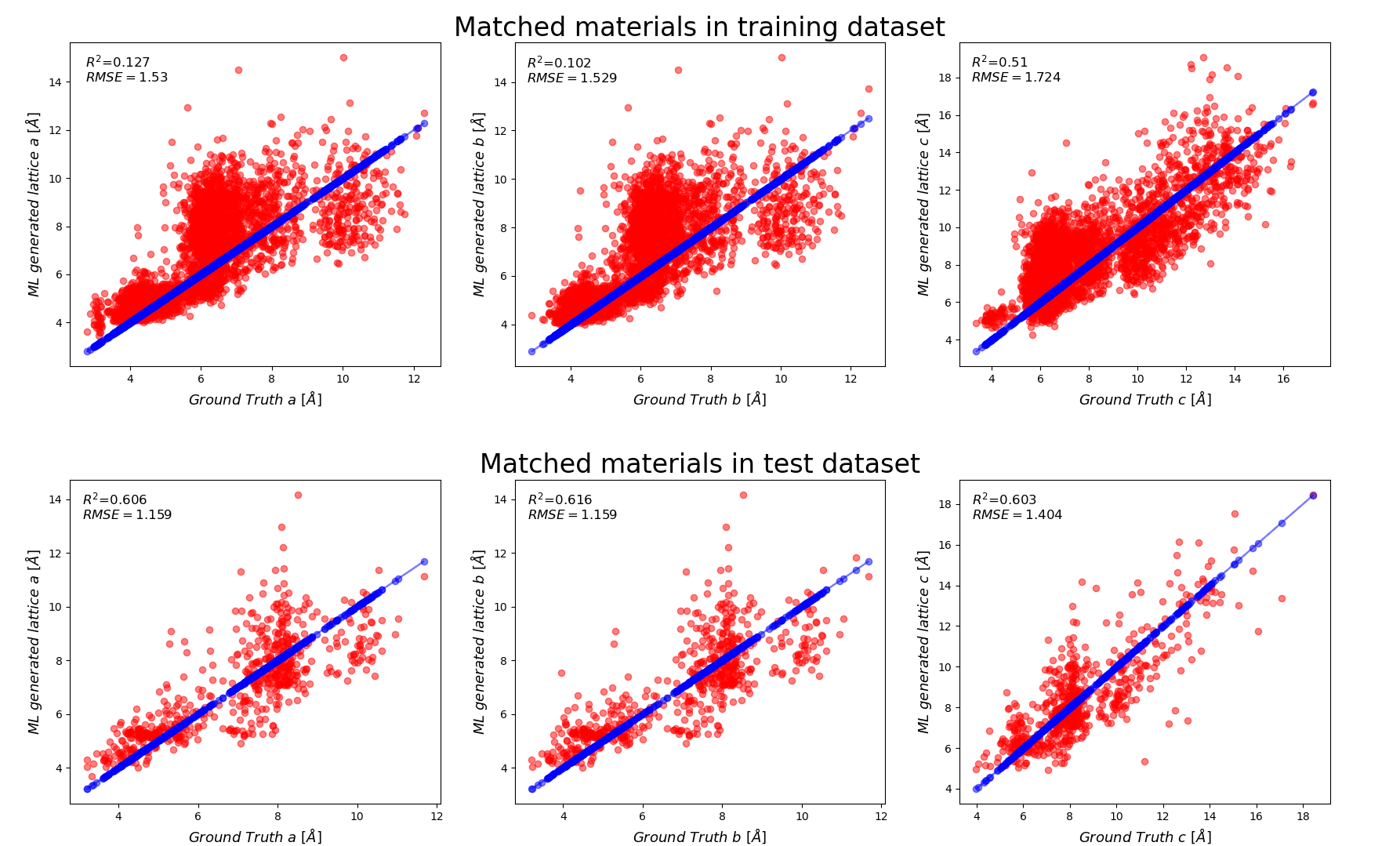}	
  	    \caption {The parity plot for lattice lengths of generated and matched materials.}
  	    \label{fig:mlat}
    \end{subfigure}
    \caption{\textbf{Analysis of rediscovered materials.} (a) The discovery rate of prototypes in MIO and TST. The number of unique prototypes is 26,463 (42,072) in \textbf{MIO} and the number of unique prototypes is 8,103 (9,441) in \textbf{TST}. (b) The parity plots for lattice lengths of generated materials (red dots) and corresponding ground truth materials that match the generated materials in \textbf{MIO} and \textbf{TST} datasets. Top row is for the training dataset and bottom row is for the test dataset (OQMD v1.5), respectively. The blue dots represent the results generated by a perfect generator that rediscover all training and testing samples. $\mathrm{R^\mathit{2}}$ and $\mathrm{RMSE}$ are also used to evaluate the performance of generated lattice lengths compared to existing ones.}
    \label{fig:redis}
\end{figure}

\paragraph{Analysis of rediscovering materials in training and test datasets}
It would be helpful to show how fast our model can rediscover materials in training datase \textbf{MIO} and test dataset \textbf{TST}. To do this, we sample different number of materials and then calculate the percentage of materials rediscovered in generated materials. "Reduced Formula - Space Group ID - \# of Atoms" is defined as prototype to identify unique materials in the existing and generated materials. Figure~\ref{fig:rdcrate} shows the change of unique crystals and rediscovery rate over size of sampling materials. We start to sample materials from half million and the number ends at 60 million eventually. It is found that the percentage of unique materials (cyan line) are decreasing and gradually tend to grow flat as number of sampling materials increases. The rediscover rate of \textbf{MIO} (orange bars) increases consistently over the sampling process and it soars quickly to 42.6\% when 35 million materials are sampled. Starting from 42.6\%, the percentage of {rediscovered} materials in \textbf{MIO} grows smoothly and it reaches 52.0\% when the sampling size is 60 million. Similar growing patterns can be observed for the rediscover rate for the test dataset, as shown by blue line in Figure~\ref{fig:rdcrate}. The rediscover rate reaches 43\% at the end of 60 million sampling size. This percentage is lower than that of training dataset because of the different proportions over the 20 space groups in training and test datasets as shown in Supplementary Table 1 and 2.

After rediscovering the materials in \textbf{MIO} and \textbf{TST} from the generated materials when sampling size is 60 million, we utilize \texttt{StructureMatcher} from \texttt{pymatgen}~\cite{ong2013python} to test whether the generated materials match the rediscovered materials and to calculate RMS displacement between two matched structures considering all invariances of materials. Because one prototype might correspond to multiple structures in existing and generated materials, we only show the least RMS displacement by exhausting each pair of existing and generated materials for this prototype. The match rate is the percentage of materials satisfying the criteria \texttt{ltol=0.2, stol=0.3, angle\_tol=5.0}. The match rate and RMS are 25.4\% and 0.05 for training dataset and are 17.7\% and 0.085 for test dataset, respectively. Figure~\ref{fig:mlat} shows parity plot that compares generated lattice lengths against DFT calculated lattice lengths. Surprisingly, the co-relation between the discovered materials in test dataset and generated materials is better than in training dataset in terms of $\mathrm{R^\mathit{2}}$ . The $\mathrm{R^\mathit{2}}$ for lattice a, b, and c in test dataset are 0.606, 0.616, and 0.606, respectively as in Figure~\ref{fig:mlat}, which increases $\mathrm{R^\mathit{2}}$ as in training dataset by a factor of six except for lattice c. The rediscovered materials in training dataset have larger lattice a and b and we find that these materials mostly are with cubic space groups. It seems that our approach tends to generate more realistic lattice for non-cubic space groups than cubic space groups in rediscovered materials.







\paragraph{DFT validation}

\begin{figure}[H]
    \centering
    \captionsetup{justification=centering}
    \begin{subfigure}{0.5\textwidth}
        \centering
        \includegraphics[width=0.85\linewidth]{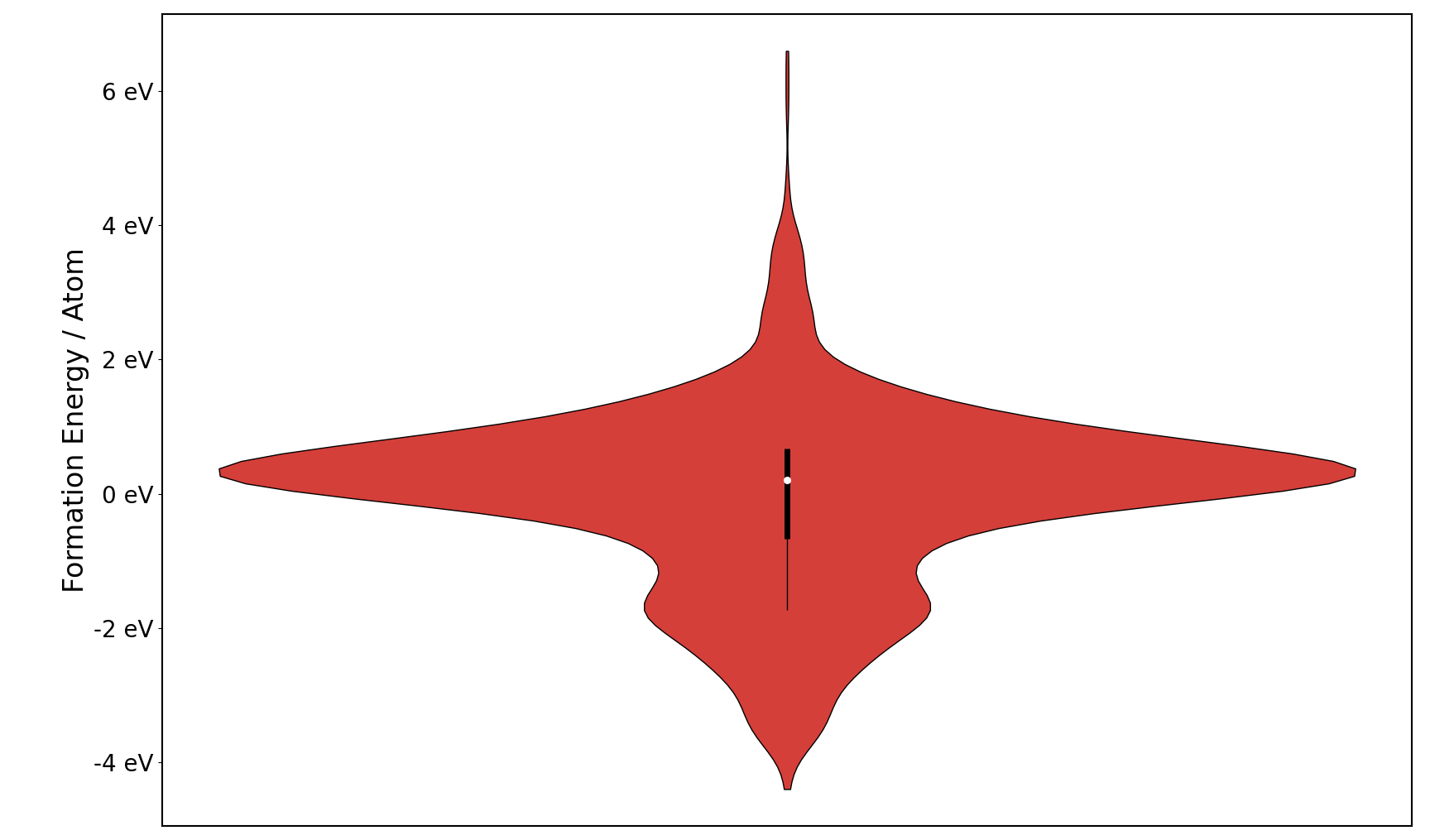}	
      	\caption{Formation energy.}
      	\label{adx-fig:fe_violin}
    \end{subfigure}\hfill
    \begin{subfigure}{0.5\textwidth}
        \centering
        \includegraphics[width=0.85\linewidth]{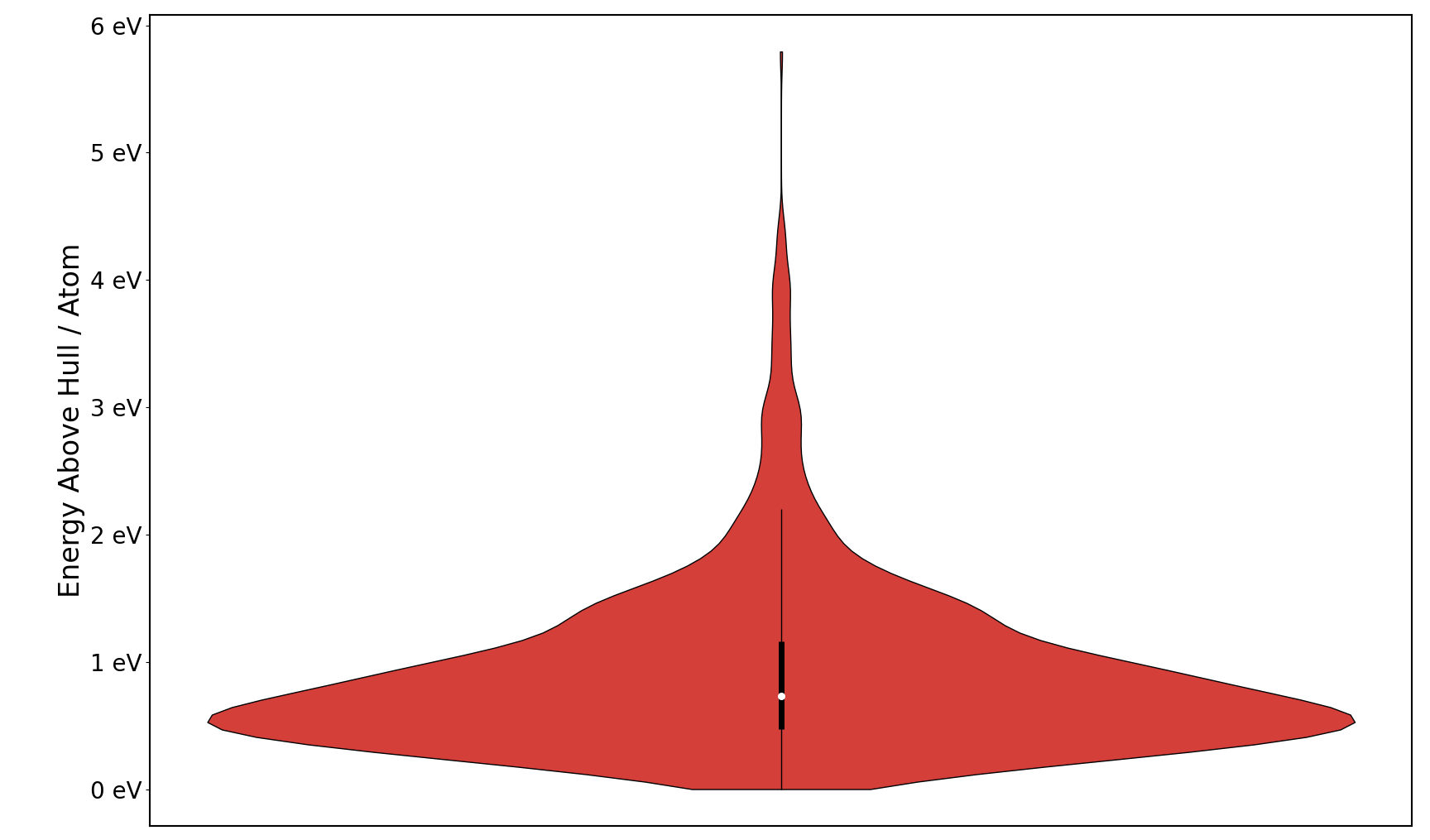}	
  	    \caption {Energy above hull.}
  	    \label{adx-fig:ehull}
    \end{subfigure}\hfill
    \caption{\textbf{The distribution of formation energy for 1,863 materials and energy above hull for 1,579 materials.} (a) 39.6\% materials are with negative formation energy. (b) Three materials are with energy above hull equal to zero and 106 ones with energy above hull less than 0.25 eV/atom among 1579 materials. }
    \label{fig:energy-mat}
\end{figure}

\begin{figure}[H]
\centering
\begin{subfigure}[b]{0.75\textwidth}
   \includegraphics[width=1.0\linewidth]{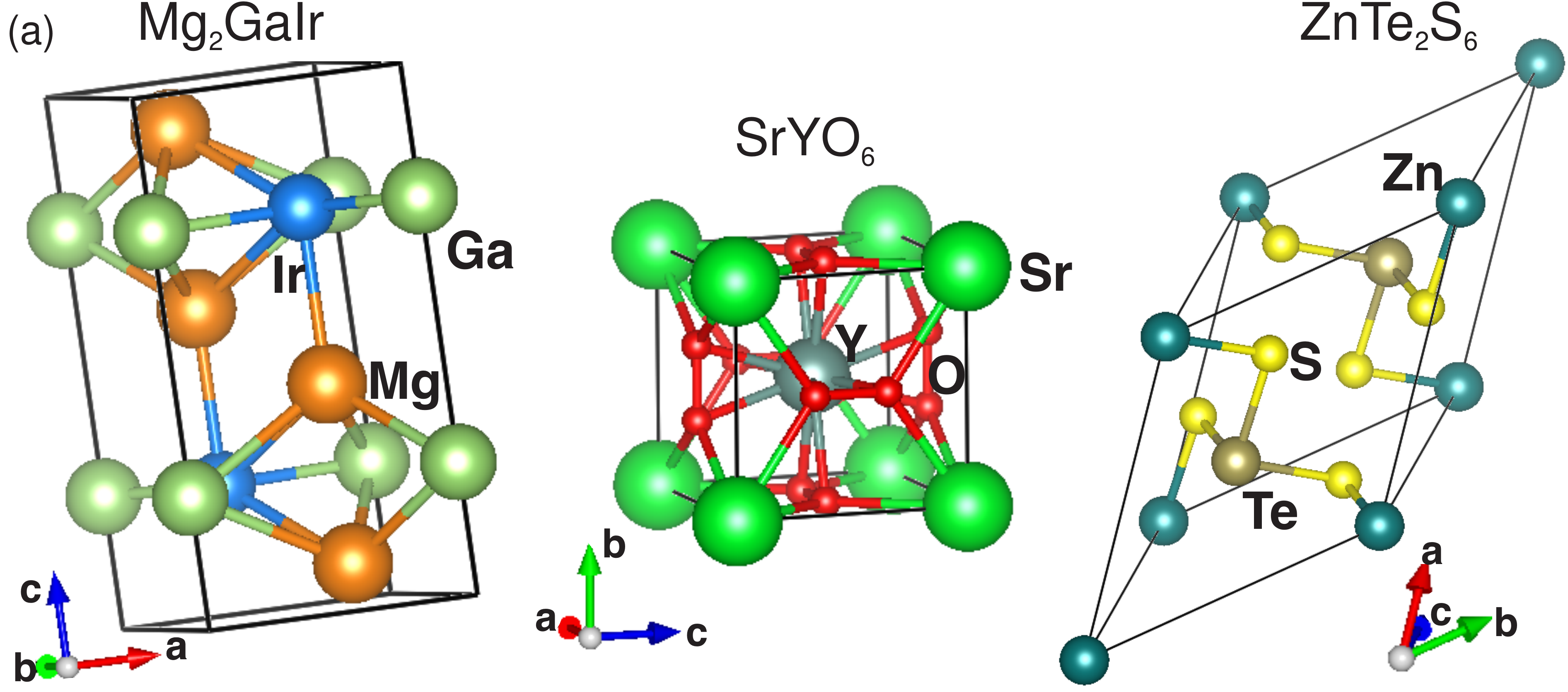}
\end{subfigure}

\begin{subfigure}[b]{0.75\textwidth}
   \includegraphics[width=1\linewidth]{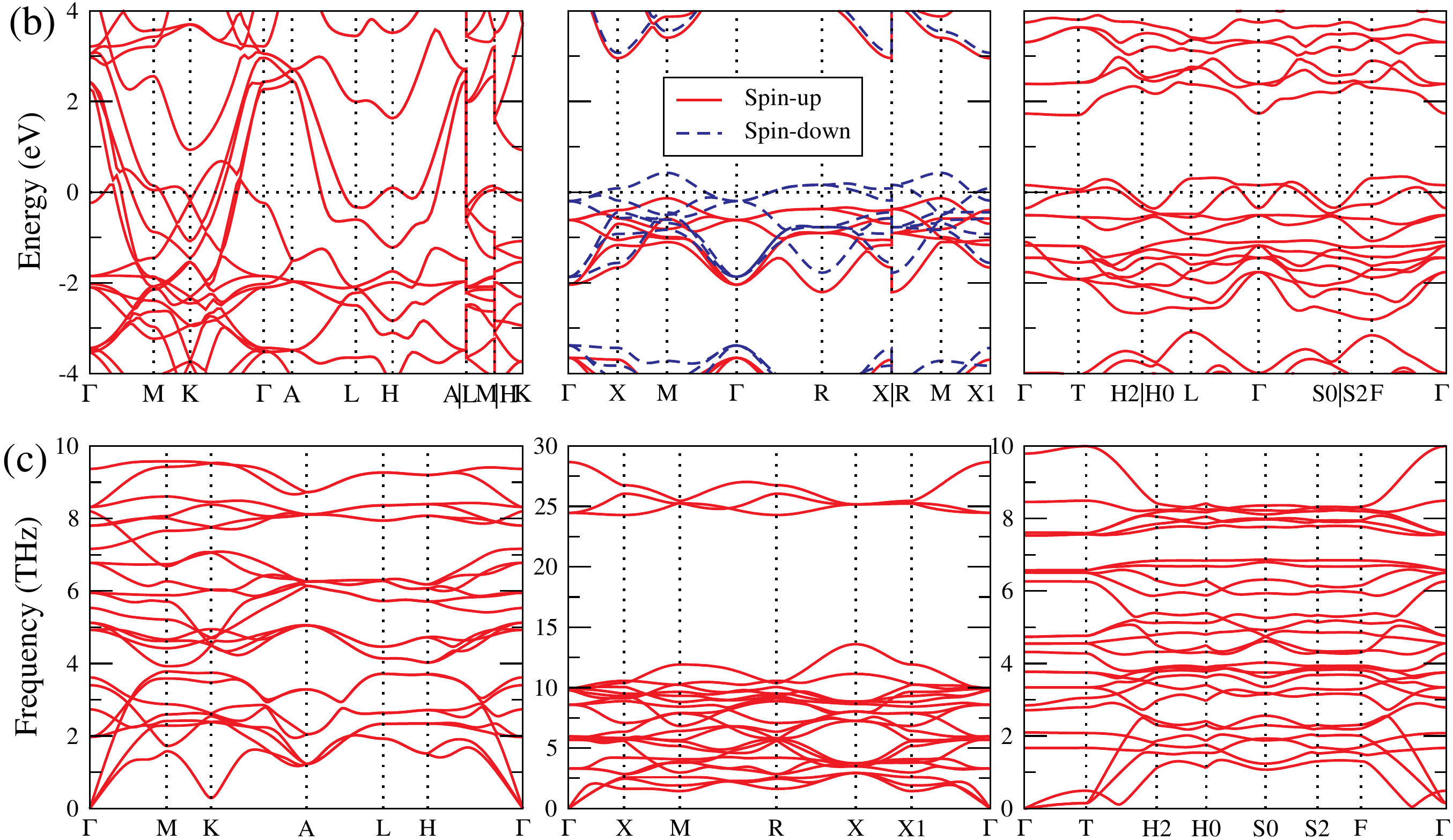}
\end{subfigure}
\caption{\textbf{The structures of Mg$_2$GaIr, SrYO$_6$ and ZnTe$_2$S$_6$ and their corresponding DFT calculated properties.} The (a) structures, (b) electronic band structures, and (c) phonon dispersion of the stable materials.}
\label{fig:struct-en-phonons}
\end{figure}

We use the same 2,000 materials as in Bayesian optimization for DFT verification. Out of 2,000 generated crystals, 93.5\% (1869) are successfully optimized, which is significantly better than 33.8\% of CubicGAN as reported in~\cite{zhao2021high}. Figure~\ref{adx-fig:fe_violin} demonstrates the distribution of formation energy of successfully optimized materials after removing 6 materials with formation energy larger than 10 eV. It is observed that most of the materials have formation energy around 0 eV and 39.6\% of them have negative formation energy. Negative formation energy indicates potentially stable materials. Figure~\ref{adx-fig:ehull} shows the distribution of 1,579 materials that have energy-above-hull after removing one material with super large energy-above-hull (1160.8 eV). The energy-above-hulls are calculated using the Pymatgen's Phase diagram analyzer \cite{ong2013python}.
Energy above hull is a stronger indicator whether the materials are stable or not. Overall, 3 materials with energy above hull of 0 eV/atom and 106 (5.3\%) ones with energy above hull less than 0.25 eV/atom, which further indicates our model can generate reliably stable materials. All the optimized materials are included in the supplementary materials.

\begin{figure}[H]
	\centering
	\includegraphics[width=0.99\linewidth]{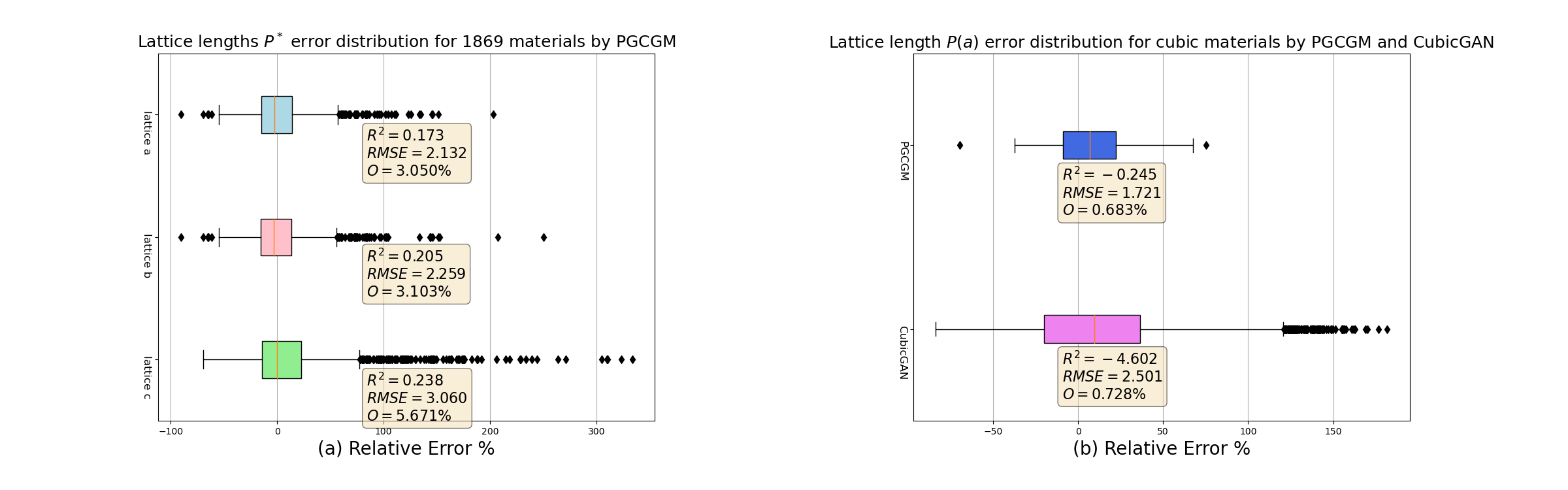}	
  	\caption {\textbf{The comparison of lattice parameters $P^*$ generation performance for 20 space groups and space groups in cubic system.} Lattice angles are constrained to fixed values by virtue of crystal systems of 20 space groups. Relative error is  calculate by $(\mathrm{length}_\mathit{GEN} - \mathrm{length}_\mathit{DFT})/\mathrm{length}_\mathit{DFT}$, where $\mathrm{length}_\mathit{GEN}$ is the generated lattice length and $\mathrm{length}_\mathit{DFT}$ is the relaxed lattice length. The error boxplot in subfigure (a) and (b)  extends from the first quartile to the third quartile of the relative error, with the line in the median. Two whiskers extend from the box by 1.5x the inter-quartile range. Outliers lie out of the whiskers. The bounding boxes correspond to each box plot above them and $\mathrm{R^\mathit{2}}$, $\mathrm{RMSE}$, and $\mathit{O}$ are used to evaluate the lattice lengths generation performance. $\mathit{O}$ means the percentage of outliers in the box plots. (a) The error distribution of three lattice lengths for 1,869 materials generated/relaxed in PGCGM. (b) The error distribution of one lattice length for cubic materials generated/relaxed in PGCGM and CubicGAN, respectively. There are 293 cubic materials optimized in PGCGM and 14,432 cubic materials optimized in CubicGAN successfully.}
  	\label{fig:lattice}
\end{figure}

Pair-wise atom distance based loss not only constrains the two atoms in a reasonable range, but also helps generate lattice lengths close to DFT-calculated ones. To demonstrate this, we calculate relative error, $\mathrm{R^\mathit{2}}$, $\mathrm{RMSE}$, and $\mathit{O}$ (outliers percentage) for lattice lengths for 1,869 materials as shown in left panel of Figure~\ref{fig:lattice} and for only 293 cubic materials by PGCGM and 14,432 cubic materials by CubicGAN as shown in right panel of Figure~\ref{fig:lattice}. In terms of relative error, we can find that the mean relative error of lattice lengths is much more close to zero regardless of when comparing 1,869 materials or just cubic materials by PGCGM with cubic materials by CubicGAN, which indicates that PGCGM tends to generated precise lattice lengths. In addition, the outliers of lattice lengths in 1,869 materials by PGCGM scatter across 100\% and cubic materials from 1,869 ones only have two outliers compared to CubicGAN whose outliers cluster near to 150\% even though CubicGAN overall has a lower outliers percentage. We also evaluate the lattice lengths generation performance between PGCGM and CubicGAN with $\mathrm{R^\mathit{2}}$ and $\mathrm{RMSE}$. For 1,869 materials, the generated lattice lengths by PGCGM better fit to the DFT calculated lattice lengths than CubicGAN in terms of $\mathrm{R^\mathit{2}}$. In terms of $\mathrm{RMSE}$, PGCGM is generally slight better CubicGAN. When only comparing cubic materials, PGCGM significantly outperforms CubicGAN in terms of both $\mathrm{R^\mathit{2}}$ and $\mathrm{RMSE}$. Although it is not a direct comparison between PGCGM and CubicGAN, all this findings indicate that our model can generate high quality materials with reasonable lattice lengths. 

In Table~\ref{tab:mats}, 20 structures with lowest formation energy are selected for 20 space groups as in the dataset. Before any post-processing, both materials has a large number of atoms. It is easily found that the atoms of the same elements are crowded together in column of \textit{GEN}. After clustering and merging the atoms of the same elements, the number of atoms drop rapidly, such as space group 164 (from 36 to 14) and space group 227 (from 432 to 32) as shown in column of \textit{MER}. Column \textit{OPT} shows the crystal structures after DFT optimization. Columns of Formula (GEN) and Fromula (MER) show the formulas before and after clustering and merging, respectively. We can find that the formulas may be changed after clustering and merging even though we only keep materials that do not change space group when conducting clustering and merging. The reason behind this is because we employ base atom sites in generated materials. Note that line 10 of Algorithm~\ref{adx-alg:coord} tends to fail because of the decimal mantissa of base atom sites  which in return can easily lead to the large number of atoms when converting from base atom sites to full atom sites via Algorithm~\ref{adx-alg:coord}, particularly space groups with large affine matrix, such as 227 and 225 as shown in Table~\ref{tab:mats} .

\begin{table}[H]
\centering
\caption{20 example optimized crystals with lowest energy for 20 space group. GEN: generated; \#: no.of atoms; MER: merged; OPT: optimized; FE: formation energy; SG: space group.}
\begin{tabular}{|c|c|c|c|c|c|c|c|c|}
\hline
Formula (GEN) & GEN & \# & Formula (MER)&MER& OPT & \# &FE(eV) &SG  \\ \hline

RbScN & \adjustimage{height=0.99cm,valign=m}{figures/G_RbScN}& 36 &RbScN12 &\adjustimage{height=0.99cm,valign=m}{figures/M_RbScN}  & \adjustimage{height=0.99cm,valign=m}{figures/O_RbScN} & 14 & -4.400& 164 \\ \hline
 
Cu4BN4 & \adjustimage{height=0.99cm,valign=m}{figures/G_Cu4BN4} & 432 &CuBN2 & \adjustimage{height=0.99cm,valign=m}{figures/M_CuBN2} & \adjustimage{height=0.99cm,valign=m}{figures/O_CuBN2} & 32 &-1.711 &227  \\ \hline

Al(BrN)2 & \adjustimage{height=0.99cm,valign=m}{figures/G_Al_BrN_2}& 36&Al(BrN4)2 &\adjustimage{height=0.99cm,valign=m}{figures/M_Al_BrN4_2}  & \adjustimage{height=0.99cm,valign=m}{figures/O_Al_BrN4_2} & 22 & -3.659& 139 \\ \hline
 
HfNF & \adjustimage{height=0.99cm,valign=m}{figures/G_HfNF} & 36&HfNF3 & \adjustimage{height=0.99cm,valign=m}{figures/M_HfNF3} & \adjustimage{height=0.99cm,valign=m}{figures/O_HfNF3} & 16 &-3.658 &186  \\ \hline

Ti2VN2 & \adjustimage{height=0.99cm,valign=m}{figures/G_Ti2VN2} & 32&TiVN8 & \adjustimage{height=0.99cm,valign=m}{figures/M_TiVN8} & \adjustimage{height=0.99cm,valign=m}{figures/O_TiVN8} & 20 &-3.503 &129  \\ \hline

HfNCl & \adjustimage{height=0.99cm,valign=m}{figures/G_HfNCl} & 60&HfN2Cl3 & \adjustimage{height=0.99cm,valign=m}{figures/M_HfN2Cl3} & \adjustimage{height=0.99cm,valign=m}{figures/O_HfN2Cl3} & 30 &-3.142 & 194 \\ \hline

HfBi2F2 & \adjustimage{height=0.99cm,valign=m}{figures/G_HfBi2F2} & 40&HfBiF8 & \adjustimage{height=0.99cm,valign=m}{figures/M_HfBiF8} & \adjustimage{height=0.99cm,valign=m}{figures/O_HfBiF8} &10  &-3.314 &123  \\ \hline

Ti2GeN2 & \adjustimage{height=0.99cm,valign=m}{figures/G_Ti2GeN2} & 120&Ti(GeN8)3 & \adjustimage{height=0.99cm,valign=m}{figures/M_Ti_GeN8_3} & \adjustimage{height=0.99cm,valign=m}{figures/O_Ti_GeN8_3} & 28 &-2.579 &221  \\ \hline

ScInN2 & \adjustimage{height=0.99cm,valign=m}{figures/G_ScInN2} &32&Sc2InN8  & \adjustimage{height=0.99cm,valign=m}{figures/M_Sc2InN8} & \adjustimage{height=0.99cm,valign=m}{figures/O_Sc2InN8} &22  &-3.690 &71  \\ \hline

Zr2VN2 & \adjustimage{height=0.99cm,valign=m}{figures/G_Zr2VN2} & 60&Zr(VN4)3 & \adjustimage{height=0.99cm,valign=m}{figures/M_Zr_VN4_3} & \adjustimage{height=0.99cm,valign=m}{figures/O_Zr_VN4_3} &16  &-3.666 &191  \\ \hline

TiNF & \adjustimage{height=0.99cm,valign=m}{figures/G_TiNF} &24&Ti(NF)2  & \adjustimage{height=0.99cm,valign=m}{figures/M_Ti_NF_2} & \adjustimage{height=0.99cm,valign=m}{figures/O_Ti_NF_2} &20  &-3.819 &62  \\ \hline

Mg4NF4 & \adjustimage{height=0.99cm,valign=m}{figures/G_Mg4NF4} &216& MgNF6 & \adjustimage{height=0.99cm,valign=m}{figures/M_MgNF6} & \adjustimage{height=0.99cm,valign=m}{figures/O_MgNF6} &32  &-2.267 &216  \\ \hline

ZrMnO & \adjustimage{height=0.99cm,valign=m}{figures/G_ZrMnO} &108&ZrMn2O6  & \adjustimage{height=0.99cm,valign=m}{figures/M_ZrMn2O6} & \adjustimage{height=0.99cm,valign=m}{figures/O_ZrMn2O6} &27  &-3.487 &166  \\ \hline

BeNO & \adjustimage{height=0.99cm,valign=m}{figures/G_BeNO} &48&BeNO  & \adjustimage{height=0.99cm,valign=m}{figures/M_BeNO} & \adjustimage{height=0.99cm,valign=m}{figures/O_BeNO} &24  &-3.721 &63  \\ \hline

Sc(HO2)2 & \adjustimage{height=0.99cm,valign=m}{figures/G_Sc_HO2_2} &56&Sc2HO4  & \adjustimage{height=0.99cm,valign=m}{figures/M_Sc2HO4} & \adjustimage{height=0.99cm,valign=m}{figures/O_Sc2HO4} &28  &-3.672 &141  \\ \hline

CaCdF & \adjustimage{height=0.99cm,valign=m}{figures/G_CaCdF} &48&Ca2CdF4  & \adjustimage{height=0.99cm,valign=m}{figures/M_Ca2CdF4} & \adjustimage{height=0.99cm,valign=m}{figures/O_Ca2CdF4} &28  &-3.494 &122  \\ \hline

Hf4Os8F & \adjustimage{height=0.99cm,valign=m}{figures/G_Hf4Os8F} &312&HfOsF6  & \adjustimage{height=0.99cm,valign=m}{figures/M_HfOsF6} & \adjustimage{height=0.99cm,valign=m}{figures/O_HfOsF6} &32  &-3.178 &225  \\ \hline

Mn2Sn2N & \adjustimage{height=0.99cm,valign=m}{figures/G_Mn2Sn2N} &80&MnSnN4  & \adjustimage{height=0.99cm,valign=m}{figures/M_MnSnN4} & \adjustimage{height=0.99cm,valign=m}{figures/O_MnSnN4} &28  &-2.611 &140  \\ \hline

ZrPoN & \adjustimage{height=0.99cm,valign=m}{figures/G_ZrPoN} &54&ZrPoN6  & \adjustimage{height=0.99cm,valign=m}{figures/M_ZrPoN6} & \adjustimage{height=0.99cm,valign=m}{figures/O_ZrPoN6} &24  &-3.945 &148  \\ \hline

MgZrO & \adjustimage{height=0.99cm,valign=m}{figures/G_MgZrO} &108&MgZrO3  & \adjustimage{height=0.99cm,valign=m}{figures/M_MgZrO3} & \adjustimage{height=0.99cm,valign=m}{figures/O_MgZrO3} &30  &-4.340 &167  \\ \hline

\end{tabular}
\label{tab:mats}
\end{table}

\paragraph{Case study of three example stable materials}

We discovered three compounds with Mg$_2$GaIr, SrYO$_6$ and ZnTe$_2$S$_6$ chemical formulas, which are thermodynamically stable with negative formation energies and zero energy above hull. The structures we found have P6\_3/mc [194] (hexagonal), Pm-3 [200] (cubic) and [148] (trigonal) space group symmetries [space group number] (crystal systems) for  Mg$_2$GaIr, SrYO$_6$ and ZnTe$_2$S$_6$ compounds, respectively. Figure \ref{fig:struct-en-phonons} (a) shows the structures of those three materials. The lattice parameters for Mg$_2$GaIr material are $\mathit{a}=\mathit{b}= 4.38$ \AA,  $\mathit{c}=8.54$ {\AA}, $\mathit{\mathbf{\beta}}=\mathit{\mathbf{\beta}}= 90^0$ and $\mathit{\mathbf{\gamma}} = 120^0$, while that for SrYO$_6$ material are $\mathit{a}=\mathit{b}=\mathit{c}= 4.61$ {\AA} and $\mathit{\mathbf{\beta}}=\mathit{\mathbf{\beta}}=\mathit{\mathbf{\gamma}} =90^0$. Moreover, the lattice parameters for ZnTe$_2$S$_6$ material are $\mathit{a}=\mathit{b}=\mathit{c} = 7.57$ {\AA} and $\mathit{\mathbf{\beta}}=\mathit{\mathbf{\beta}}=\mathit{\mathbf{\gamma}}=49.38^0$. As shown in Table~\ref{tab:elec-mag}, our spin-polarized DFT calculations show that the both Mg$_2$GaIr and ZnTe$_2$S$_6$ compounds have the non-magnetic ground states, whereas SrYO$_6$ material has a ferromagnetic ground state with a total magnetic moment of 1 $\mathit{\mu}_\mathit{B}$. Figure \ref{fig:struct-en-phonons} (b) contains the electronic band structures for each stable material. It is clear that both Mg$_2$GaIr and ZnTe$_2$S$_6$ compounds are metals. However, we can see spin-splitting in SrYO$_6$ ferromagnetic material. In this compound only spin-down electrons cross the Fermi level, while spin-up electrons have a band gap of 3.09 eV. Thus, this is a half-metal where spin-down electrons show metallic character, while spin-up electrons are insulating. Half metallicity is widely investigated for spintronics and it is vital for developing memory devices and computer processors \cite{half-metal}.

\begin{table}[H]
\centering
\caption{The electronic and magnetic properties of the stable materials. The material type (metal or semiconductor), band gap, magnetic ground state: GS (non magnetic: NM or ferromagnetic: FM) and the magnetic moment: $\mu$ are reported for each stable material. For the half-metal, the band gaps for both spin types (Up and Down) are mentioned. }
\label{tab:elec-mag}
\begin{tabular}{|c|c|c|c|c|}
\hline
Material & Material Type & Band Gap (eV) & Magnetic GS & $\mathit{\mu}$ ($\mathit{\mu}_\mathit{B}$) \\ \hline
Mg$_2$GaIr  & Metal         & 0        & NM          & 0                       \\ \hline
SrYO$_6$    & Half-metal & Up:3.09, Down:0.00   & FM          & 1                       \\ \hline
ZnTe$_2$S$_6$  & Metal         & 0        & NM          & 0                       \\ \hline
\end{tabular}
\end{table}

\begin{table}[H]
\centering
\caption{The elastic constants ($\mathit{C}_\mathit{ij}$),  bulk ($\mathit{B}$) modulus, shear ($\mathit{G}$) modulus and Young's ($\mathit{Y}$) modulus in GPa and Poisson's ratio ($\mathit{\nu}$) for the stable materials.}
\label{tab:elastic}
\begin{tabular}{|c|c|c|c|c|c|c|c|c|c|c|}
\hline
Material & $\mathit{C}_\mathit{11}$ & $\mathit{C}_\mathit{12}$ & $\mathit{C}_\mathit{33}$ & $\mathit{C}_\mathit{13}$ & $\mathit{C}_\mathit{44}$ & $\mathit{C}_\mathit{66}$ 
& $\mathit{B}$       & $\mathit{G}$      & $\mathit{Y}$       & $\mathit{\nu}$\\ \hline
Mg$_2$GaIr  & 177.32  & 98.68   & 132.77  & 52.04   & 36.276   & 36.276 
& 96.44  & 40.67 & 106.96 & 0.315  \\ \hline
SrYO$_6$    & 181.61  & 60.18   & 181.61  & 60.18   & 37.951   & 37.951 
& 100.66 & 45.85 & 119.42 & 0.302   \\ \hline
ZnTe$_2$S$_6$  & 83.13   & 31.49   & 4.56    & 0.85    & 0.382    & 25.821 
& 15.34  & 7.69  & 19.77  & 0.285\\ \hline
\end{tabular}
\end{table}

Table~\ref{tab:elastic} contains the elastic constants and the mechanical properties of the stable materials. The elastic stability criteria (Born criteria) for Mg$_2$GaIr with P6\_3/mm space group symmetry are $\mathit{C}_\mathit{11} > |\mathit{C}_\mathit{12}|$, $2*\mathit{C}_\mathit{13}^\mathit{2} < \mathit{C}_\mathit{33}(\mathit{C}_\mathit{11} + \mathit{C}_\mathit{12})$, and $\mathit{C}_\mathit{44} > 0$. The Born criteria for SrYO$_6$ with Pm-3 space group symmetry are $\mathit{C}_\mathit{11} - \mathit{C}_\mathit{12} > 0$, $\mathit{C}_\mathit{11} + 2\mathit{C}_\mathit{12} > 0$, $\mathit{C}_\mathit{44} >0$, and that for ZnTe$_2$S$_6$ with R-3 space group symmetry are $\mathit{C}_\mathit{11} > |\mathit{C}_\mathit{12}|$, $\mathit{C}_\mathit{13}^\mathit{2} < 0.5*\mathit{C}_\mathit{33}(\mathit{C}_\mathit{11} + \mathit{C}_\mathit{12})$ and $\mathit{C}_\mathit{14}^\mathit{2} + \mathit{C}_\mathit{15}^\mathit{2} < 0.5*\mathit{C}_\mathit{44}*(\mathit{C}_\mathit{11}-\mathit{C}_\mathit{12})$, and $\mathit{C}_\mathit{44} >0$ \cite{Meachanical_Stability}. It is clear that all those three materials comply with their elastic stability criteria implying they are mechanically stable. Table~\ref{tab:elastic} also has the bulk ($\mathit{B}$) modulus, shear ($\mathit{G}$) modulus and Young's ($\mathit{Y}$) modulus and Poisson's ratio ($\mathit{\nu}$) for Mg$_2$GaIr, SrYO$_6$, and ZnTe$_2$S$_6$ compounds. We used Hill approximation as implemented in vaspkit code \cite{Vaspkit,hill1952elastic}. It is clear that Mg$_2$GaIr and  SrYO$_6$ materials have approximately same $B$, $G$, and $Y$ values, while those values for ZnTe$_2$S$_6$ compound is considerably lower. In Table~\ref{tab:elastic}, $\mathit{\nu}$ of Mg$_2$GaIr has the highest value, while lowest $\mathit{\nu}$ can be obtained from  ZnTe$_2$S$_6$. Furthermore, we used Phonopy code \cite{Phonopy} to calculate the phonon dispersion relations for the above materials. As shown in Fig.~\ref{fig:struct-en-phonons} (c), there are no imaginary phonon modes (negative frequencies) indicating those three materials are dynamically stable at 0 K.




In this work, we propose a physics guided deep crystal generative model (PGCGM), in which two kinds of physics based losses are invented in the generator to improve the quality of generated materials. The atom distance based losses constrain the atom distance in a certain range in the generated materials and thus the generated lattice parameters fall into reasonable range too. To fulfill the symmetry requirements, the model transforms the implicit rules between base atoms sites and full atom sites into explicit cost functions. Two baseline methods are compared and PGCGM achieves the best performance across all evaluation metrics. In particular, PGCGM significantly outperforms the two baseline models in terms of property distribution metric which is a much stronger indicator to show the reality of the generated materials~\cite{xie2021crystal}. In addition, we use BOWSR to optimize 2,000 randomly selected materials in each method. Our approach has the best match rate calculated between the Generative model-generated materials and BOWSR-optimized materials, which further demonstrate our method can generate realistic materials.

In order to see how our approach can rediscover materials in existing databases, we sample different size of materials and calculate rediscover rate for training and test datasets. We can observe a clear trend of increased rediscover rate over sampling size. There is no clear saturation point of rediscover rate at the end of 60 million sampled materials as in CubicGAN~\cite{zhao2021high}. The reasons are: 1) the possible design space of 20 space groups (5 crystal systems) in this work are much bigger than 3 space groups (only cubic crystal systems) in CubicGAN; 2) CubicGAN uses special fractional coordinates while PGCGM generates fractional coordinates in full space, which means PGCGM has a significantly broader space to explore new materials. Furthermore, 1,869 out of 2,000 materials are successfully optimized by DFT calculation. Among 1,869 materials, 39.6\% possess negative formation energy and 5.3\% with energy above hull less than 0.25 eV/atom, indicating that invented physics guided losses help generate stable crystal structures effectively. This research gives a deep insight into how physics losses help generate realistic materials and offers an approach to expand the diversity of generated materials.

Due to the difficulty to generate 500,000 structures by the diffusion based generative model CDVAE, we conduct a small-scale evaluation and comparison. We generated 1,100 structures using CDVAE, among which 78.2\% (860) are pymatgen readable. We then analyzed the space groups of these 860 structures and find out of the top 10 space groups with the most structures (a total of 790), 97.8\% (773) structures have low symmetry with space group number less or equal to 25 and 582 (73.7\%) of them have no symmetry. However, it is shown~\cite{shao2022symmetry} that high-symmetry materials tend to form stable structures and have larger potential to have good functional properties. It seems that like other VAE based generator models, CDVAE also has difficulty in generating high-symmetry structures as PGCGM does.

\section*{Methods}

\subsection*{Problem statement and notations}

Our data-driven generative models of crystal structures are first trained with known crystal structures in materials databases. Since crystal materials are periodic structures, instead of representing the infinite structures, the structure of an inorganic material is represented by a unit cell in material science, which is the smallest unit that completely reflects the arrangement of atoms in the 3D space. Given a generated unit cell, it can be used to build the periodic structure of an inorganic material by repeating it multiple or infinite times along three directions to form a super cell. A material $\mathcal{M}$ can then be denoted as following:

\begin{equation}
    \mathcal{M} = (\mathbf{E}, \mathbf{B}, \mathbf{P}, \mathbf{O})
\end{equation}
where

(a) $\mathbf{E}=(\mathit{e}_\mathit{0}, \mathit{e}_\mathit{1}, \mathit{e}_\mathit{2})\in\mathbb{E}$ denotes elements in materials, where $\mathbb{E}$ is the element set in periodic table. In this work, we only deal with ternary materials so that there are only three unique elements in the unit cell; 

(b) $\mathbf{B}=(\mathbf{b}_\mathit{0},\mathbf{b}_\mathit{1},\mathbf{b}_\mathit{2} )\in \mathbb{R}^{\mathit{3}\times \mathit{3}}$ denotes the base atom sites, which are the symmetry equivalent positions. termed . $\mathbf{b}_\mathit{i}$ is fractional coordinates of an atom denoted by $[\mathit{u},\mathit{v},\mathit{w}]^\mathit{T}$. We choose materials that one element only has one base atom site so that three atom sites can be used to represent the atom positions. Moreover, any one atom site of each element can be considered as the base atom site for that element; 

(c) $\mathbf{P}=(\mathit{a},\mathit{b},\mathit{c},\mathit{\mathbf{\beta}},\mathit{\mathbf{\beta}}, \mathit{\mathbf{\gamma}})\in \mathbb{R}$ are six lattice parameters that define three lengths and three angles of the unit cell; 

(d) $\mathbf{O}=( \mathbf{t}_\mathit{0}, \mathbf{t}_\mathit{1}, \ldots, \mathbf{t}_\mathit{n})\in\mathbb{R}^{\mathit{n}\times \mathit{4}\times \mathit{4}}$ denotes affine matrix that represents the symmetry operations defined by space groups $\mathrm{sgp}$. $\mathbf{t}_\mathit{j}$ is one affine operator containing the rotation and translation matrices. $\mathit{n}$ is determined by space groups. Generally the higher symmetry of a space group, the larger $\mathit{n}$. $\mathit{n}$ can be as small as 1 or as large as 192.

Now we can model the generation of materials as follows:

\begin{equation}\label{eq:gen}
    (\mathbf{B}, \mathbf{P}) =f_\mathit{\theta}(\mathbf{Z}, \mathbf{E}, \mathrm{sgp}),
\end{equation}
where $f_\mathit{\theta} $ is the generative model that learns the knowledge of forming crystal structures given inputs of random noise $\mathbf{Z}$, element set $\mathbf{E}$, and space group $\mathrm{sgp}$.

\subsection*{Overall architecture of physics guided crystal generative model}
\begin{figure*}[ht]
	\centering
	\includegraphics[width=\linewidth]{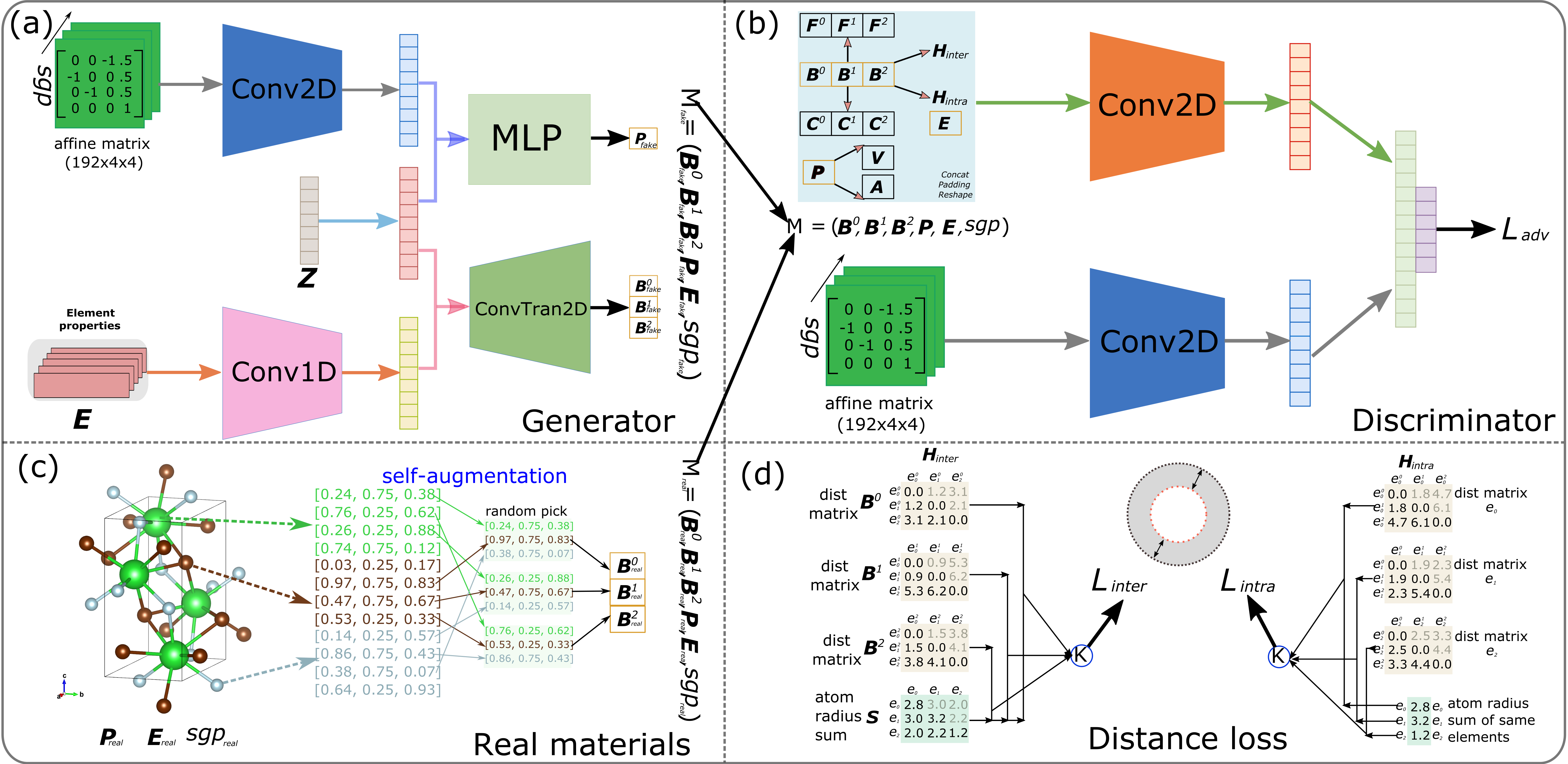}
  	\caption {\textbf{The overview of physics guided crystal generative model (PGCGM).} The PGCGM comprises four components. (a) The generator takes affine matrix $\mathbf{O}$, random noise $\mathbf{Z}$, and element properties $\mathbf{E}$ as inputs. The affine matrix and random noise are projected to two vectors by 2D convolutional networks and fully connected layers, respectively and then the two vectors are merged and projected to generate lattice parameters $\mathbf{P}^{*}$ by fully connected layers. The element properties are projected to a vector by 1D convolutional networks and then it is merged with the vector projected from random noise to generate three sets of base atom sites $(\mathbf{B}_\mathit{fake}^\mathit{0},\mathbf{B}_\mathit{fake}^\mathit{1},\mathbf{B}_\mathit{fake}^\mathit{2})$. (b) The discriminator has two input branches. It shares with the same affine matrix branch as in the generator. The assembled crystal representation matrix from three sets of base atom sites, lattice parameters, and properties calculated from them is used as the input to 2D convolutional networks. The assembled matrix is zero-padded to form a matrix with shape of $3\times8\times8$. (c) The self-augmentation performed on the base atom sites. We choose three sets of base atom sites from three elements randomly and with space group, we can calculate more crystal information to assemble the input matrix for the discriminator. (d) Inter- and intra-atom distance matrices ($\mathbf{H}_\mathit{inter}$ and $\mathbf{H}_\mathit{inter}$) are calculated from three sets of base atom sites for both real and fake materials. Then we design distance based losses to constrain the distance between two atoms in a certain range as shown in the grey area form by two circles. }\label{fig:mainframe}
\end{figure*}

Our Physics Guided Crystal Generative Model (PGCGM) is shown in Figure~\ref{fig:mainframe}. The PGCGM mainly consists of four major components: (1) discriminator, (2) generator, (3) self-augmentation, and (4) atom distance matrix/loss calculation module. In the generator and discriminator, affine matrix is integrated into the training to generate fake materials and tell fake materials from real ones, respectively. Affine matrix is related to symmetry information for space groups. The implicit combination of affine matrix and base atom sites can help keep the symmetry when generating materials. Self-augmentation increases the training materials for underrepresented space groups by randomly forming base atom sites. With three sets of base atom sites, we can not only have a fixed size of input to the discriminator, but also deduce more physical information for crystals to help the discriminator better distinguish real materials from fake ones. Furthermore, we design two kinds of physics guided losses. Any set of base atom sites can be converted to full set of unique atom sites. When generating three sets of base atom sites, it implicitly can be stated that the three sets of base atom sites should be different but the full atom sites converted from them separately ought to be same. Hence a specific loss is invented to explicitly incorporate this rule into training of the generator. In order to restrict the two atoms in the 3D space to be not too close or not too distant, inter- and intra-atom distance losses are proposed. With distance loss, the generator further can generate reasonable lattice parameters in order to push any pair of atoms to fall into a certain range. 

\paragraph{Discriminator}
There are two input branches for crystal representation and affine matrix in Discriminator as in sub-figure (b) of Figure~\ref{fig:mainframe}. Each branch is forwarded to a 2D convolutional block and the learnt features are concatenated together. The concatenated vector is sent to a couple of fully connected layers to get the discriminative score. We have three different sets of base atom sites in our inputs and with the affine matrix branch, it helps to implicitly learn the knowledge of how affine matrix transforms base atom sites into full atom sites. The detailed architectures of two convolutional blocks can be found in Table S3 in the supplementary materials.

\paragraph{Generator}
The architecture of generator is shown in sub-figure (a) of Figure~\ref{fig:mainframe},which contains three branches. Conditioning on element constituents and space group, the generator outputs three sets of base atom sites \\$(\mathbf{B}_\mathit{fake}^\mathit{0},\mathbf{B}_\mathit{fake}^\mathit{1},\mathbf{B}_\mathit{fake}^\mathit{2})$ and unit cell length $\mathbf{P}^{*}$. Then we re-formalize Equation~\eqref{eq:gen} as follow:
\begin{equation}\label{eq:rgen}
    (\mathbf{B}_\mathit{fake}^\mathit{0},\mathbf{B}_\mathit{fake}^\mathit{1},\mathbf{B}_\mathit{fake}^\mathit{2}, \mathbf{P}_\mathit{fake}^{*}) =f_\mathit{\theta}^{*}(\mathbf{Z}, \mathbf{E}, \mathrm{sgp}).
\end{equation}
Taking random noise $\mathbf{Z}$, space group $\mathrm{sgp}$, and element properties matrix $\mathbf{E}$ as inputs, the generator can generate a material with the same lattice parameters and space group but different representations of the base atom sites when merely sampling one material. Our goal here is that the generated three sets of base atom sites belong to the same material. Random noise $\mathbf{Z}$ is mapped to a dense vector a fully connected layer. The space group branch is the same as in discriminator. Element matrix $\mathbf{E}$ is forwarded to a 1D convolutional layer (Conv1D). The outputs of random noise and space group branches are combined together as the inputs to a multi-layer perceptron (MLP) block to generate unit cell length $\mathbf{P}_\mathit{fake}^{*}$. The outputs of random noise and element branches are combined together as the inputs to 2D deconvolutional layers (ConvTran2D) to generate three sets of base atom sites $(\mathbf{B}_\mathit{fake}^\mathit{0},\mathbf{B}_\mathit{fake}^\mathit{1},\mathbf{B}_\mathit{fake}^\mathit{2})$. The detailed descriptions for MLP, Conv1D, and ConvTran2D can be found in Supplementary Table 4.

\paragraph{Physics guided loss function}

The original GAN~\cite{goodfellow2014generative} is notoriously hard to train because of saturation and mode collapse in discriminator. We take advantage of WGAN-GP~\cite{gulrajani2017improved} with gradient penalty to enhance the training stability in our network. WGAN-GP changes the Sigmoid function of the discriminator to a 1-Lipschitz function while introducing a gradients penalty term to enforce the norm of gradients to be close to 1. The loss function is described in Equation~\eqref{eq:adv_loss}:
\begin{equation}\label{eq:adv_loss}
      \begin{aligned}
        \hat{\mathcal{M}^{*}} &= \mathit{\epsilon}\mathcal{M}^{*}_\mathit{real}+(1-\mathit{\epsilon})\mathcal{M}^{*}_\mathit{fake},~~~\mathit{\epsilon} \sim U(0,1),\\ 
        \mathcal{L}_\mathit{dis} &= D(\mathcal{M}^{*}_\mathit{fake}) - D(\mathcal{M}^{*}_\mathit{real}) + \mathit{\lambda}_\mathit{d}(\norm{\nabla_{\hat{\mathcal{M}^{*}}}D(\hat{\mathcal{M}^{*}})}_\mathit{2}-1)^\mathit{2},\\
        \mathcal{L}_\mathit{adv} &= -D(\mathcal{M}^{*}_\mathit{fake}),
      \end{aligned}
\end{equation}
where $\hat{\mathcal{M}^{*}}$ is linearly interpolated between real {materials $\mathcal{M}^{*}_\mathit{real}$} and fake materials {$\mathcal{M}^{*}_\mathit{fake}$} and $\mathit{\epsilon}$ is uniformly sampled from 0 and 1. $\mathcal{L}_\mathit{dis}$ and $\mathcal{L}_\mathit{adv}$ represent the loss function of the discriminator and adversarial loss for generator respectively. The third term in $\mathcal{L}_\mathit{dis}$ is the gradient penalty and $\mathit{\lambda}_\mathit{d}$ is set {to} 10. $D(.)$ means the score result from the discriminator.

\textit{Atom Distance Losses.} To ensure that the atoms in generated crystal structures are not crowded or not too far apart from each other, we introduce the inter- and intra-atom distance based losses as following:

\begin{equation}
  \begin{aligned}
    \mathcal{L}_\mathit{inter} &= \frac{1}{\mathrm{9N}}\sum_{\mathit{i}=1}^\mathit{N}\{[\max(\mathbf{H}_\mathit{inter}, \mathit{\phi}_\mathit{inter}^\mathit{upper} \mathbf{S}_\mathit{inter})-\mathit{\phi}_\mathit{inter}^\mathit{upper} \mathbf{S}_\mathit{inter}]^\mathit{2}\\
    &+[\min(\mathbf{H}_\mathit{inter}, \mathit{\phi}_\mathit{inter}^\mathit{lower} \mathbf{S}_\mathit{inter})-\mathit{\phi}_\mathit{inter}^\mathit{lower} \mathbf{S}_\mathit{inter}]^\mathit{2}\}, \\
    \mathcal{L}_\mathit{inter} &= \frac{1}{\mathrm{9N}}\sum_{\mathit{i}=1}^\mathit{N}\{[\max(\mathbf{H}_\mathit{inter}, \mathit{\phi}_\mathit{inter}^\mathit{upper} \mathbf{S}_\mathit{inter})-\mathit{\phi}_\mathit{inter}^\mathit{upper} \mathbf{S}_\mathit{inter}]^\mathit{2}\\
    &+[\min(\mathbf{H}_\mathit{inter}, \mathit{\phi}_\mathit{inter}^\mathit{lower} \mathbf{S}_\mathit{inter})-\mathit{\phi}_\mathit{inter}^\mathit{lower} \mathbf{S}_\mathit{inter}]^\mathit{2}\},
  \end{aligned}
\end{equation}
 where $\mathcal{L}_\mathit{inter}$ constrains the distance in $\mathbf{H}_\mathit{inter}$, which describes inter-atom distance matrices. $[\max(\mathbf{H}_\mathit{inter}, \mathit{\phi}_\mathit{inter}^\mathit{upper} \mathbf{S}_\mathit{inter})-\mathit{\phi}_\mathit{inter}^\mathit{upper} \mathbf{S}_\mathit{inter}]^\mathit{2}$ enforces the atom distance to be smaller than $\mathit{\phi}_\mathit{inter}^\mathit{upper} \mathbf{S}_\mathit{inter}$ and $[\min(\mathbf{H}_\mathit{inter}, \mathit{\phi}_\mathit{inter}^\mathit{lower} \mathbf{S}_\mathit{inter})-\mathit{\phi}_\mathit{inter}^\mathit{lower} \mathbf{S}_\mathit{inter}]^\mathit{2}$ enforces the atom distance to be bigger than $\mathit{\phi}_\mathit{inter}^\mathit{lower} \mathbf{S}_\mathit{inter}$. $\mathbf{S}_\mathit{inter}$ are atom radius sum corresponding to each pair of atoms in $\mathbf{H}_\mathit{inter}$ and $\mathit{\phi}_\mathit{inter}^\mathit{upper}$ and $\mathit{\phi}_\mathit{inter}^\mathit{lower}$ are control weights for upper and lower bound of inter-atom distance, respectively. In this way, the distance of two atoms is constrained to be in the grey area indicated by two circles in sub-figure (d) of Figure~\ref{fig:mainframe}. Similarly, $\mathcal{L}_\mathit{inter}$ constrains the distance in a range in $\mathbf{H}_\mathit{inter}$ which describes intra-atom distance matrices. {$\mathbf{S}_\mathit{inter}$ are atom radius sum corresponding to each pair of atoms in $\mathbf{H}_\mathit{inter}$ and $\mathit{\phi}_\mathit{inter}^\mathit{upper}$ and $\mathit{\phi}_\mathit{inter}^\mathit{lower}$ are control weights for upper and lower bound of inter-atom distance, respectively.} $\mathit{N}$ is batch size and 9 is the number of distance value in $\mathbf{H}_\mathit{inter}$ and $\mathbf{H}_\mathit{inter}$.

\textit{Symmetry-compliant Base and Average Full Coordinates Losses.} The generator generates three sets of base atom sites \\$(\mathbf{B}_\mathit{fake}^\mathit{0},\mathbf{B}_\mathit{fake}^\mathit{1},\mathbf{B}_\mathit{fake}^\mathit{2})$ which are used to generate the full site coordinates using the symmetric operation defined by the space group. The averaged transformation to $(\mathbf{F}_\mathit{fake}^\mathit{0},\mathbf{F}_\mathit{fake}^\mathit{1},\mathbf{F}_\mathit{fake}^\mathit{2})$ from base atom sites should be exactly same. With these implicit rules, we design two losses to explicitly enforce them in the generator as expressed below: 

\begin{equation}
    \resizebox{.9\hsize}{!}{$%
    \begin{aligned}
    \mathcal{L}_{full} = \frac{1}{\mathrm{9N}}\sum_{\mathit{i}=1}^\mathit{N}\{&\max(0,cos(\frac{\mathbf{F}_\mathit{fake}^\mathit{0}}{\norm{\mathbf{F}_\mathit{fake}^\mathit{0}}_\mathit{2}}, \frac{\mathbf{F}_\mathit{fake}^\mathit{1}}{\norm{\mathbf{F}_\mathit{fake}^\mathit{1}}_\mathit{2}}))&\mathcal{L}_\mathit{base} = \frac{1}{\mathrm{9N}}\sum_{\mathit{i}=1}^\mathit{N}\{&(1-cos(\frac{\mathbf{B}_\mathit{fake}^\mathit{0}}{\norm{\mathbf{B}_\mathit{fake}^\mathit{0}}_\mathit{2}}, \frac{\mathbf{B}_\mathit{fake}^\mathit{1}}{\norm{\mathbf{B}_\mathit{fake}^\mathit{1}}_\mathit{2}}))\\
    +&\max(0,cos(\frac{\mathbf{F}_\mathit{fake}^\mathit{1}}{\norm{\mathbf{F}_\mathit{fake}^\mathit{1}}_\mathit{2}}, \frac{\mathbf{F}_\mathit{fake}^\mathit{2}}{\norm{\mathbf{F}_\mathit{fake}^\mathit{2}}_\mathit{2}}))& +&(1-cos(\frac{\mathbf{B}_\mathit{fake}^\mathit{1}}{\norm{\mathbf{B}_\mathit{fake}^\mathit{1}}_\mathit{2}}, \frac{\mathbf{B}_\mathit{fake}^\mathit{2}}{\norm{\mathbf{B}_\mathit{fake}^\mathit{2}}_\mathit{2}}))\\
    +&\max(0,cos(\frac{\mathbf{F}_\mathit{fake}^\mathit{0}}{\norm{\mathbf{F}_\mathit{fake}^\mathit{0}}_\mathit{2}}, \frac{\mathbf{F}_\mathit{fake}^\mathit{2}}{\norm{\mathbf{F}_\mathit{fake}^\mathit{2}}_\mathit{2}}))\},& +&(1-cos(\frac{\mathbf{B}_\mathit{fake}^\mathit{0}}{\norm{\mathbf{B}_\mathit{fake}^\mathit{0}}_\mathit{2}}, \frac{\mathbf{B}_\mathit{fake}^\mathit{2}}{\norm{\mathbf{B}_\mathit{fake}^\mathit{2}}_\mathit{2}}))\},
    \end{aligned}
    $%
  }
\end{equation}

where $cos$ is cosine similarity function. We normalize each coordinate value across the mini-batch of size $N$. 9 is the number of coordinates.

\textit{Full Generator Loss.} By combining above losses, we can achieve our full loss for the generator:
\begin{equation}
    \mathcal{L}_{gen} = \mathcal{L}_\mathit{adv} +\mathit{\lambda}_\mathit{1}\mathcal{L}_\mathit{inter} + \mathit{\lambda}_\mathit{2}\mathcal{L}_\mathit{inter}+ \mathit{\lambda}_\mathit{3}\mathcal{L}_{full}+ \mathit{\lambda}_\mathit{4}\mathcal{L}_\mathit{base}
\end{equation}
{where $\mathit{\lambda}_\mathit{1}$, $\mathit{\lambda}_\mathit{2}$, $\mathit{\lambda}_\mathit{3}$, and $\mathit{\lambda}_\mathit{4}$ are balancing parameters. }

\paragraph{Crystal Symmetry based Self-augmentation}
Data augmentation is commonly used for images in which operations such as rotation of an image does not change its label. Similarly, self-augmentation as we define here is used to do data augmentation based on the symmetry-oriented Wyckoff position representation of CIF files. In the representation of symmetric crystals, the coordinates of the non-equivalent positions (Wyckoff positions) are just one of a set of possible positions as defined by the symmetric operations of the space group. So, for each structure file, we can use the set of symmetric operations of the space group to transform the Wyckoff position coordinates without changing the structure, which can then generate more equivalent structure samples. The generation of atom coordinates that meet the symmetry constraints is one of the most challenging tasks in crystal generation. In order to make the fixed size of representation for crystals (details in Table~\ref{tab:sym-input}), we use base atom sites. As shown in sub-figure (c) of Figure~\ref{fig:mainframe}, we can use any atom site of each element to form a set of base atom sites. Instead of randomly selecting them, we choose three atoms for three elements individually using steps as shown in below:
\begin{enumerate}[labelsep=0ex,align=left,start=2]
    \item[Step~1.~] Randomly {select} the first element $\mathit{e}_\mathit{0}$ and one atom position $\mathbf{b}_\mathit{0}$ for it;
    \item[Step~2.~] Randomly {select} the second element $\mathit{e}_\mathit{1}$ from the rest two elements and find the closest atom $\mathbf{b}_\mathit{1}$ {to atom $\mathbf{b}_\mathit{0}$} in the first step;
    \item[Step~3.~] Calculate the atom distance from the atoms of the last element $\mathit{e}_\mathit{2}$ to {the atom $\mathbf{b}_\mathit{0}$ and the atom $\mathbf{b}_\mathit{1}$} respectively, then sum the atom distance element-wise and the atom of the last element with the smallest sum is considered as the closest atom $\mathbf{b}_\mathit{2}$ to the selected atoms in the first and second steps;
   \item[Step~4.~] Repeat Steps 2, 3, and 4 three times to obtain three sets of base atom sites $(\mathbf{B}_\mathit{real}^\mathit{0}, \mathbf{B}_\mathit{real}^\mathit{1}, \mathbf{B}_\mathit{real}^\mathit{2})$;
    \item[Step~5.~] Repeat last five steps 31 times.
\end{enumerate}

In Step 4, we use three sets of base atom sites as part of inputs to the discriminator so that we can obtain more information from crystal structures. In this work, we obtain three sets of base atom sites 32 times repeatedly as in Step 5.

\paragraph{Atom Clustering and Merging}{For crystals with high symmetry, the number of atoms in the unit cell tends to be very large after conversion by Algorithm 1. We propose a post-processing method to reduce the number of atoms by clustering and merging. Firstly, we cluster the nearby atoms of the same elements by forming flat clusters from hierarchical clustering~\cite{mullner2011modern,bar2001fast}. The maximum atom distance allowed in our research is 1.2 times the atom radius sum. Secondly, we merge the atoms in the same clusters considering periodic attributes of crystal structures.}

\subsection*{Implementation details}

\paragraph{Materials Representation}
We use $\mathcal{M} = (\mathbf{E}, \mathbf{B}, \mathbf{P}, \mathbf{O})$ to completely describe a crystal material. As shown in mainframe of PGCGM, however, we use three sets of base atom sites $(\mathbf{B}^\mathit{0},\mathbf{B}^\mathit{1},\mathbf{B}^\mathit{2})$. Thus here we re-formulate a material as $\mathcal{M}^{*} = (\mathbf{B}^\mathit{0},\mathbf{B}^\mathit{1},\mathbf{B}^\mathit{2}, \mathbf{P}, \mathbf{E}, \mathrm{sgp})$. The space group $\mathrm{sgp}$ is used to link to the affine matrix $\mathbf{O}$. We can use $(\mathbf{B}^\mathit{0},\mathbf{B}^\mathit{1},\mathbf{B}^\mathit{2})$ in $\mathcal{M}^{*}$ to calculate physical properties as inputs to the discriminator and to design physics-based losses. Three sets of base atom sites are useful for two reasons: (1) we want to add more crystal information for the discriminator and let the discriminator have enough information to tell real materials from fake ones; (2) With more base atom sites, we can calculate more atom distances as the physical constraints in the generator and the inputs to the discriminator.

\textit{All Fractional Coordinates }{We use affine matrix $\mathbf{O}$ to acquire the whole atom sites in the unit cell as shown in Algorithm 1. Since the number of affine operators in $\mathbf{O}$ varies in space groups, we zero-pad the affine matrices as large as $192\times 4\times 4$. We then transform each base atom site by the affine matrix and get a coordinates matrix $\mathbf{F}_\mathit{all}$ with shape of $192\times 3\times 3$. Affine transformation leads to duplicate fractional coordinates. In material science, practitioners usually remove the duplicates. However, uniqueness calculation is not differentiable and it requires lots of time to do it. We choose to average along with the first dimension of $\mathbf{F}_\mathit{all}$ to get three sets of averaged full fractional coordinates $(\mathbf{F}^\mathit{0}, \mathbf{F}^\mathit{1}, \mathbf{F}^\mathit{2})$, each of which is with shape of $3\times 3$.

For a real material, base atom sites $(\mathbf{B}_\mathit{real}^\mathit{0},\mathbf{B}_\mathit{real}^\mathit{1},\mathbf{B}_\mathit{real}^\mathit{2})$ can be transformed into the same average full fractional coordinates, which means $\mathbf{F}_\mathit{real}^\mathit{0} = \mathbf{F}_\mathit{real}^\mathit{1} = \mathbf{F}_\mathit{real}^\mathit{2}$. When generating a fake material, base atom sites $(\mathbf{B}_\mathit{fake}^\mathit{0},\mathbf{B}_\mathit{fake}^\mathit{1},\mathbf{B}_\mathit{fake}^\mathit{2})$ are supposed to belong to the same fake material, which hopefully results in $\mathbf{F}_\mathit{fake}^\mathit{0} = \mathbf{F}_\mathit{fake}^\mathit{1} = \mathbf{F}_\mathit{fake}^\mathit{2}$. However, the transformation of $(\mathbf{B}_\mathit{fake}^\mathit{0},\mathbf{B}_\mathit{fake}^\mathit{1},\mathbf{B}_\mathit{fake}^\mathit{2})$ might slightly deviate from the goal. Thus using $(\mathbf{F}^\mathit{0}, \mathbf{F}^\mathit{1}, \mathbf{F}^\mathit{2})$ in real and fake materials implicitly adds physical constraints, which somehow pushes the generator to generate different sets of base atom sites for a same material, which increases chances to generate good materials in return.

}

\textit{Base Cartesian Coordinates }{Three sets of Cartesian coordinates can be calculated for each set of base atom sites by Equation~\eqref{eq:f2c} and we denote them by $(\mathbf{C}^\mathit{0}, \mathbf{C}^\mathit{1}, \mathbf{C}^\mathit{2})$.
}

\textit{Atom Distance Matrices }{Given three sets of base atom sites $(\mathbf{B}^\mathit{0},\mathbf{B}^\mathit{1},\mathbf{B}^\mathit{2})$, we calculate the atom distance matrices $\mathbf{H}_\mathit{inter}$ and $\mathbf{H}_\mathit{inter}$ as shown in sub-figure (d) of Figure~\ref{fig:mainframe}. We firstly calculate pair-wise different atom distance matrix for each base atom site $\mathbf{B}^\mathit{j},\  \mathit{j}=0,1,2$ and return only values in upper triangle of corresponding distance matrix termed by $\mathbf{H}_\mathit{inter}$. 
Then we select three atoms belonging to the same element to form a set of three atom sites for three elements and calculate pair-wise same atom distance matrix and again return only values in upper triangle of corresponding distance matrix termed by $\mathbf{H}_\mathit{inter}$. The final shape of $\mathbf{H}_\mathit{inter}$ and $\mathbf{H}_\mathit{inter}$ both is $3\times 3$.
}

\textit{Lattice Parameters }{The volume of the unit cell can be calculated by lattice parameters $\mathbf{P}$. We repeat the scalar volume three times to get the volume vector $\mathbf{V}$. We also use the lattice matrix $\mathbf{A}$ in Equation~\eqref{eq:lm} as part of the inputs to the discriminator.
}

\textit{Element Properties }{We use 23 properties as shown in Table~\ref{tab:elem-prop} to formalize element matrix $\mathbf{E}$.}

\begin{table}[H]
    \centering
    \caption{3 element properties used for element embedding in PGCGM}
    \begin{tabular}{|c|c|c|}
    \hline
     \textbf{Properties}& \textbf{Properties} & \textbf{Properties} \\ \hline
     Atomic number&Average ionic radius  & noble gas or not \\ \hline
     Pauling electronegativity&Average cationic radius  & transition metal or not \\ \hline
     Periodic table row&Average anionic radius  & post transition metal or not \\ \hline
     Periodic table group&Sum of all ionic radii  & metalloid or not \\ \hline
     Atomic mass& Maximum oxidation state & alkali or not \\ \hline
    Atomic radius & Minimum oxidation state & alkaline or not \\ \hline
     Mendeleev number&Average all common oxidation states  & halogen or not \\ \hline
     Molar volume& Average all known oxidation states &  \\ \hline
    \end{tabular}
    \label{tab:elem-prop}
\end{table}

Now we list all parts of inputs to the discriminator in Table~\ref{tab:sym-input}. $\mathbf{P}^{*}$ only contains the lengths because the angles are either $(90^{\circ}, 90^{\circ}, 90^{\circ})$ or $(90^{\circ}, 90^{\circ}, 120^{\circ})$ in the training materials. Thus instead of generating three angles in $\mathbf{P}$ for fake materials, we build a map between angles and the space group $\mathrm{sgp}$. Then we concatenate all parts and a zero matrix of shape $3\times 3$ into a matrix of shape of $3\times 64$. The matrix is finally reshaped into $3\times 8\times 8$ as the inputs to the discriminator.

\begin{table}[H]
\centering
\caption{Symbols and their shape used in inputs to the discriminator.}
\begin{tabular}{|c|c|c|c|}
\hline
 symbol&shape  &symbol &shape   \\ \hline
 $(\mathbf{B}^\mathit{0},\mathbf{B}^\mathit{1},\mathbf{B}^\mathit{2})$& $3\times 9$ & $\mathbf{P}^{*}$ & $3\times 1$  \\ \hline
 $(\mathbf{F}^\mathit{0},\mathbf{F}^\mathit{1},\mathbf{F}^\mathit{2})$& $3\times 9$ & $\mathbf{V}$ &  $3\times 1$  \\ \hline
 $(\mathbf{C}^\mathit{0},\mathbf{C}^\mathit{1},\mathbf{C}^\mathit{2})$& $3\times 9$ & $(\mathbf{H}_\mathit{inter}, \mathbf{H}_\mathit{inter})$ & $3\times 6 $  \\ \hline
  $\mathbf{E}$& $3\times 23 $ & $\mathbf{A}$ &  $3\times 3 $  \\ \hline
\end{tabular}
\label{tab:sym-input}
\end{table}

\paragraph{Mathematical conversion in crystal representations}Fractional coordinates can be converted to Cartesian coordinates $[\mathit{x},\mathit{y},\mathit{y}]^{T}$ using~\cite{wiki-fc}:

\begin{equation}\label{eq:f2c}
    \begin{bmatrix}
    \mathit{x} \\
    \mathit{y} \\
    \mathit{z} \\
  \end{bmatrix} = \mathbf{A} \cdot \begin{bmatrix}
    \mathit{u} \\
    \mathit{v} \\
    \mathit{w} \\
  \end{bmatrix},
\end{equation}

where $\mathbf{A}$ is a lattice matrix calculated by lattice parameters $\mathbf{P}$ using:

\begin{equation}\label{eq:lm}
    \mathbf{A}= \begin{bmatrix}
    \mathit{a}&\mathit{b}\cos{\mathbf{\gamma}}&\mathit{c}\cos{\mathbf{\beta}} \\
    0&\mathit{b}\sin{\mathbf{\gamma}}&\mathit{c}\frac{\cos{\mathbf{\beta}}-\cos{\mathbf{\beta}}\cos{\mathbf{\gamma}} }{\sin{\mathbf{\gamma}}} \\
    0&0&\frac{\mathit{V}}{\mathit{a}\mathit{b}\sin{\mathbf{\gamma}}} \\
  \end{bmatrix},
\end{equation}
where $\mathit{V}=\mathit{a}\mathit{b}\mathit{c}\sqrt{1-\cos^\mathit{2}{\mathbf{\beta}}-\cos^\mathit{2}{\mathbf{\beta}}-\cos^\mathit{2}{\mathbf{\gamma}} +2\cos{\mathbf{\beta}}\cos{\mathbf{\beta}}\cos{\mathbf{\gamma}}  }$ is the volume of the unit cell.

In order to acquire all atom positions in the unit cell, each base atom site can be converted by affine matrix $\mathbf{O}$. The conversion procedure is summarized in Algorithm~\ref{adx-alg:coord}. Different materials vary from the number of atoms and the number of elements. In order to make a fixed size of inputs, we only use ternary materials in this research. After conversion shown in Algorithm~\ref{adx-alg:coord}, the number of atom (sites) also differs from materials. That is the reason why base atom sites (one element one base site) are used to represent atom positions. In addition, it should be noted that the calculation of the uniqueness at line 10 of Algorithm~\ref{adx-alg:coord} is not differentiable and time-consuming.

\begin{algorithm}[H]
\caption{Generate unique coordinates using base sites () and affine matrix}\label{adx-alg:coord}
\begin{algorithmic}[1]
\Require The space group $\mathrm{sgp}$, the base atom sites $\mathbf{B}$

\State $ \mathbf{O} \gets GetAffineMatrices(\mathrm{sgp})$ 
\State $\mathit{n} \gets len(\mathbf{O})$
\State $ \mathbf{coords} \gets an~empty~list$
\For{$\mathit{i} \gets 1$ to $3$}
    \State add $0$ to $\mathbf{b}_\mathit{i}$
    \State $ \mathbf{uniq} \gets an~empty~list$
    \For{$\mathit{j} \gets 1$ to $\mathit{n}$}
        \State $ \mathbf{c} \gets \mathbf{b}_\mathit{i}\cdot \mathbf{t}_\mathit{j} - \floor*{\mathbf{b}_\mathit{i}\cdot \mathbf{t}_\mathit{j}}$ 
        \State pop last element from $\mathbf{c}$
        \If{$ \mathbf{c}\ not\ in\ \mathbf{uniq} $}
            \State $add\ \mathbf{c}\ to\ \mathbf{uniq} $
        \EndIf
    \EndFor
    \State $ add\ \mathbf{uniq}\ to\ \mathbf{coords} $
\EndFor
\State \Return $\mathbf{coords}$

\end{algorithmic}
\end{algorithm}

\subsection*{Evaluation Metrics}
Past studies in crystal generation used different evaluation metrics, making it hard to compare different methods. Here, we create a set of metrics to evaluate our method and two baselines. 1) \textit{Validity\cite{xie2021crystal}.} Following~\cite{court20203}, we consider a crystal structure as valid when the shortest distance between any two atoms is bigger than 0.5\r{A}. Following CubicGAN, we calculate the overall charge of a crystal structure using \texttt{pymatgen}~\cite{ong2013python} and if it is neutral, then it is valid. Also, we count the number of structures after post-processing in our method and we apply the same post-processing onto the CubicGAN. 2) \textit{Property distribution \cite{xie2021crystal}} We calculate wasserstein distance (WD) between the property distribution of generated materials and materials in test dataset \textbf{TST}. The properties we used are minimum atom distance, maximum atom distance, and density. 3) \textit{Diversity.} We calculate the diversity of compositions, which means the ratio of unique number of compositions in generated structures. 4) DFT validation of generated structures. We find that both the pair-wise atomic distance and property distribution based performance metrics are indirect weak criteria in crystal structure generation as the major challenge in such models is to generate stable structures. It is thus critical to check the success rate of DFT-based structural relaxation, the thermodynamic stability (e.g. evaluated by DFT phonon dispersion calculation), and synthesizability (e.g. based on energy-above-hull calculation). 

\section*{Data Availability statement}
The raw crystal dataset is downloaded from \url{http://www.materialsproject.org}.
\section*{Code Availability statement}
The source code to generate crystals can be obtained from github at \url{https://github.com/MilesZhao/PGCGM}.

\section*{Acknowledgements}
The research reported in this work was supported in part by National Science Foundation under the grant and 2110033, 1940099 and 1905775. The views, perspectives, and content do not necessarily represent the official views of the NSF. 

\section*{Author contributions}
Conceptualization, J.H., M.H., Y.Z.; methodology,Y.Z., J.H., E.S., Z.W., M.H, N.F., M.A.; software, Y.Z., Z.W.; resources, J.H.; writing--original draft preparation, Y.Z., J.H., E.S., N.F.; writing--review and editing, J.H, E.S., M.H.; visualization, Y.Z., E.S., J.H.; supervision, J.H., M.H.;  funding acquisition, J.H. and M.H.

\section*{Competing interests}
The authors declare that they have no competing interests.

\newpage
\bibliographystyle{unsrt}
\bibliography{refs}

\begin{thebibliography}{10}

\bibitem{belsky2002new}
Alec Belsky, Mariette Hellenbrandt, Vicky~Lynn Karen, and Peter Luksch.
\newblock New developments in the inorganic crystal structure database (icsd):
  accessibility in support of materials research and design.
\newblock {\em Acta. Crystallogr. B Struct. Sci. Cryst. Eng. Mater.},
  58(3):364--369, 2002.

\bibitem{noh2019inverse}
Juhwan Noh, Jaehoon Kim, Helge~S Stein, Benjamin Sanchez-Lengeling, John~M
  Gregoire, Alan Aspuru-Guzik, and Yousung Jung.
\newblock Inverse design of solid-state materials via a continuous
  representation.
\newblock {\em Matter}, 1(5):1370--1384, 2019.

\bibitem{pyzer2015high}
Edward~O Pyzer-Knapp, Changwon Suh, Rafael G{\'o}mez-Bombarelli, Jorge
  Aguilera-Iparraguirre, and Al{\'a}n Aspuru-Guzik.
\newblock What is high-throughput virtual screening? a perspective from organic
  materials discovery.
\newblock {\em Annu. Rev. Mater. Res.}, 45:195--216, 2015.

\bibitem{jain2013commentary}
Anubhav Jain~et al.
\newblock Commentary: The materials project: A materials genome approach to
  accelerating materials innovation.
\newblock {\em APL Mater.}, 1(1):011002, 2013.

\bibitem{kirklin2015open}
Scott Kirklin~et al.
\newblock The open quantum materials database (oqmd): assessing the accuracy of
  dft formation energies.
\newblock {\em NPJ Comput. Mater.}, 1(1):1--15, 2015.

\bibitem{franceschetti1999inverse}
Alberto Franceschetti and Alex Zunger.
\newblock The inverse band-structure problem of finding an atomic configuration
  with given electronic properties.
\newblock {\em Nature}, 402(6757):60--63, 1999.

\bibitem{doll2008structure}
K~Doll, JC~Sch{\"o}n, and M~Jansen.
\newblock Structure prediction based on ab initio simulated annealing for boron
  nitride.
\newblock {\em Phys. Rev. B}, 78(14):144110, 2008.

\bibitem{amsler2010crystal}
Maximilian Amsler and Stefan Goedecker.
\newblock Crystal structure prediction using the minima hopping method.
\newblock {\em J. Chem. Phys.}, 133(22):224104, 2010.

\bibitem{flores2020crystal}
Jos{\'e}~A Flores-Livas.
\newblock Crystal structure prediction of magnetic materials.
\newblock {\em J. Phys. Condens. Matter}, 32(29):294002, 2020.

\bibitem{glass2006uspex}
Colin~W Glass, Artem~R Oganov, and Nikolaus Hansen.
\newblock Uspex—evolutionary crystal structure prediction.
\newblock {\em Comput. Phys. Commun.}, 175(11-12):713--720, 2006.

\bibitem{wang2012calypso}
Yanchao Wang, Jian Lv, Li~Zhu, and Yanming Ma.
\newblock Calypso: A method for crystal structure prediction.
\newblock {\em Comput. Phys. Commun.}, 183(10):2063--2070, 2012.

\bibitem{zhao2021high}
Yong Zhao, Mohammed Al-Fahdi, Ming Hu, Edirisuriya~MD Siriwardane, Yuqi Song,
  Alireza Nasiri, and Jianjun Hu.
\newblock High-throughput discovery of novel cubic crystal materials using deep
  generative neural networks.
\newblock {\em Advanced Science}, 8(20):2100566, 2021.

\bibitem{fuhr2022deep}
Addis~S Fuhr and Bobby~G Sumpter.
\newblock Deep generative models for materials discovery and machine
  learning-accelerated innovation.
\newblock {\em Front. in Mater.}, page 182, 2022.

\bibitem{schwalbe2020generative}
Daniel Schwalbe-Koda and Rafael G{\'o}mez-Bombarelli.
\newblock Generative models for automatic chemical design.
\newblock In {\em Machine Learning Meets Quantum Physics}, pages 445--467.
  Springer, 2020.

\bibitem{kingma2013auto}
Diederik~P Kingma and Max Welling.
\newblock Auto-encoding variational bayes.
\newblock {\em arXiv preprint arXiv:1312.6114}, 2013.

\bibitem{goodfellow2014generative}
Ian~Goodfellow et~al.
\newblock Generative adversarial nets.
\newblock {\em Adv. Neural Inf. Process. Syst.}, 27, 2014.

\bibitem{hoffmann2019data}
Jordan Hoffmann, Louis Maestrati, Yoshihide Sawada, Jian Tang, Jean~Michel
  Sellier, and Yoshua Bengio.
\newblock Data-driven approach to encoding and decoding 3-d crystal structures.
\newblock {\em arXiv preprint arXiv:1909.00949}, 2019.

\bibitem{court20203}
Callum~J Court, Batuhan Yildirim, Apoorv Jain, and Jacqueline~M Cole.
\newblock 3-d inorganic crystal structure generation and property prediction
  via representation learning.
\newblock {\em J. Chem. Inf. Model.}, 60(10):4518--4535, 2020.

\bibitem{ren2020inverse}
Zekun Ren~et al.
\newblock Inverse design of crystals using generalized invertible
  crystallographic representation.
\newblock {\em arXiv preprint arXiv:2005.07609}, 2020.

\bibitem{xie2021crystal}
Tian Xie, Xiang Fu, Octavian-Eugen Ganea, Regina Barzilay, and Tommi Jaakkola.
\newblock Crystal diffusion variational autoencoder for periodic material
  generation.
\newblock {\em arXiv preprint arXiv:2110.06197}, 2021.

\bibitem{ho2020denoising}
Jonathan Ho, Ajay Jain, and Pieter Abbeel.
\newblock Denoising diffusion probabilistic models.
\newblock {\em Adv. Neural Inf. Process. Syst.}, 33:6840--6851, 2020.

\bibitem{nouira2019crystalgan}
Asma Nouira, Nataliya Sokolovska, and Jean-Claude Crivello.
\newblock Crystalgan: Learning to discover crystallographic structures with
  generative adversarial networks.
\newblock In {\em AAAI Spring Symposium: Combining Machine Learning with
  Knowledge Engineering}, 2019.

\bibitem{zhu2017unpaired}
Jun-Yan Zhu, Taesung Park, Phillip Isola, and Alexei~A Efros.
\newblock Unpaired image-to-image translation using cycle-consistent
  adversarial networks.
\newblock In {\em Proceedings of the IEEE international conference on computer
  vision}, pages 2223--2232, 2017.

\bibitem{kim2020generative}
Sungwon Kim, Juhwan Noh, Geun~Ho Gu, Alan Aspuru-Guzik, and Yousung Jung.
\newblock Generative adversarial networks for crystal structure prediction.
\newblock {\em ACS Cent. Sci.}, 6(8):1412--1420, 2020.

\bibitem{long2021constrained}
Teng Long~et al.
\newblock Constrained crystals deep convolutional generative adversarial
  network for the inverse design of crystal structures.
\newblock 7(1):1--7, 2021.

\bibitem{ahmad2022free}
Rasool Ahmad and Wei Cai.
\newblock Free energy calculation of crystalline solids using normalizing
  flows.
\newblock {\em Model. Simul. Mat. Sci. Eng.}, 30(6):065007, 2022.

\bibitem{wirnsberger2022normalizing}
Peter Wirnsberger et alPython~materials genomics.
\newblock Normalizing flows for atomic solids.
\newblock {\em Mach. Lear.: Sci. Tech.}, 3(2):025009, 2022.

\bibitem{baird2022xtal2png}
Sterling G~Baird et~al.
\newblock xtal2png: A python package for representing crystal structure as png
  files.
\newblock {\em J. Open Source Softw.}, 7(76):4528, 2022.

\bibitem{dan2020generative}
Yabo Dan, Yong Zhao, Xiang Li, Shaobo Li, Ming Hu, and Jianjun Hu.
\newblock Generative adversarial networks (gan) based efficient sampling of
  chemical composition space for inverse design of inorganic materials.
\newblock {\em NPJ Comput. Mater.}, 6(1):1--7, 2020.

\bibitem{sawada2019study}
Yoshihide Sawada, Koji Morikawa, and Mikiya Fujii.
\newblock Study of deep generative models for inorganic chemical compositions.
\newblock {\em arXiv preprint arXiv:1910.11499}, 2019.

\bibitem{gulrajani2017improved}
Ishaan Gulrajani, Faruk Ahmed, Martin Arjovsky, Vincent Dumoulin, and Aaron
  Courville.
\newblock Improved training of wasserstein gans.
\newblock {\em arXiv preprint arXiv:1704.00028}, 2017.

\bibitem{ong2013python}
Shyue~Ping Ong~et al.
\newblock Python materials genomics (pymatgen): A robust, open-source python
  library for materials analysis.
\newblock {\em Comput. Mater. Sci.}, 68:314--319, 2013.

\bibitem{zuo2021accelerating}
Yunxing Zuo~et al.
\newblock Accelerating materials discovery with bayesian optimization and graph
  deep learning.
\newblock {\em Mater. Today}, 51:126--135, 2021.

\bibitem{xie2018crystal}
Tian Xie and Jeffrey~C Grossman.
\newblock Crystal graph convolutional neural networks for an accurate and
  interpretable prediction of material properties.
\newblock {\em Phys. Rev. Lett.}, 120(14):145301, 2018.

\bibitem{chen2019graph}
Chi Chen, Weike Ye, Yunxing Zuo, Chen Zheng, and Shyue~Ping Ong.
\newblock Graph networks as a universal machine learning framework for
  molecules and crystals.
\newblock {\em Chem. Mater.}, 31(9):3564--3572, 2019.

\bibitem{half-metal}
Warren~E. Pickett and Jagadeesh~S. Moodera.
\newblock Half metallic magnets.
\newblock {\em Phys Today}, 54(5):39--44, 2001.

\bibitem{Meachanical_Stability}
F{\'e}lix Mouhat and Fran{\c{c}}ois-Xavier Coudert.
\newblock Necessary and sufficient elastic stability conditions in various
  crystal systems.
\newblock {\em Phys. Rev. B}, 90(22):224104, 2014.

\bibitem{Vaspkit}
Vei Wang, Nan Xu, Jin-Cheng Liu, Gang Tang, and Wen-Tong Geng.
\newblock Vaspkit: A user-friendly interface facilitating high-throughput
  computing and analysis using vasp code.
\newblock {\em Comput. Phys. Commun.}, 267:108033, Oct 2021.

\bibitem{hill1952elastic}
Richard Hill.
\newblock The elastic behaviour of a crystalline aggregate.
\newblock {\em Proc. Phys. Soc. A}, 65(5):349, 1952.

\bibitem{Phonopy}
A~Togo and I~Tanaka.
\newblock First principles phonon calculations in materials science.
\newblock {\em Scr. Mater.}, 108:1--5, Nov 2015.

\bibitem{shao2022symmetry}
Xuecheng Shao~et al.
\newblock A symmetry-orientated divide-and-conquer method for crystal structure
  prediction.
\newblock {\em J. Chem. Phys.}, 156(1):014105, 2022.

\bibitem{mullner2011modern}
Daniel M{\"u}llner.
\newblock Modern hierarchical, agglomerative clustering algorithms.
\newblock {\em arXiv preprint arXiv:1109.2378}, 2011.

\bibitem{bar2001fast}
Ziv Bar-Joseph, David~K Gifford, and Tommi~S Jaakkola.
\newblock Fast optimal leaf ordering for hierarchical clustering.
\newblock {\em Bioinformatics}, 17(suppl\_1):S22--S29, 2001.

\bibitem{wiki-fc}
{Fractional coordinates}.
\newblock Fractional coordinates --- {W}ikipedia{,} the free encyclopedia,
  2021.
\newblock [Online; accessed 11-November-2021].

\end{thebibliography}

\end{document}


\section*{}
\paragraph{Supplementary file1: Physics Guided Deep Learning for Generative Design of Crystal Materials with Symmetry Constraints}

\hfill \break

Yong Zhao, Edirisuriya M. Dilanga Siriwardane, Zhenyao Wu, Nihang Fu, Mohammed Al-Fahdi, Ming Hu*, Jianjun Hu*

\setcounter{table}{0}
\renewcommand{\thetable}{S\arabic{table}}

\setcounter{figure}{0}
\renewcommand{\thefigure}{S\arabic{figure}}

\section{Dataset Curation}\label{adx:A}
We select the material data from three databases: MP~\cite{jain2013commentary}, ICSD~\cite{belsky2002new}, OQMD~\cite{kirklin2015open}. The selection criteria are described following:
\begin{enumerate}[nolistsep]
    \item Ternary materials with only three base atom sites (a.k.a. one element is allowed to have only one base atom site);
    \item Only keep materials that do not contain elements in Lanthanoid and Actinoid;
    \item Ternary materials whose space group has more than 400 materials totally in three databases;
    \item Ternary materials in OQMD whose fractional coordinates does not all belong to the set $[0.0, 0.25, 0.5, 0.75]$ since materials with fractional coordinates all falling in that set dominate the database~\cite{zhao2021high}. 
\end{enumerate}
In total, 42072 materials are selected and 20 space groups are found in those materials following above criteria. The statistics of materials in each space group is shown in Table~\ref{adx-tab:mio}.

\begingroup
\setlength{\tabcolsep}{8pt} 
\renewcommand{\arraystretch}{1.2} 
\begin{table}[ht]
\centering
\begin{tabular}{|c|c|c|c|c|c|}
\hline
SG& SG Id &\# samples & SG & SG Id &\# samples\\ \hline
$P4/mmm$ & 123 &1180   &$Immm$ &71 &4679  \\ \hline
$Fm\bar{3}m$ &225 &3716  &$Cmcm$ &63 &1004  \\ \hline
$I4_1/amd$ &141&588  &$I\bar{4}2d$ &122 &749  \\ \hline
$Pm\bar{3}m$ &221&1462  &$R\bar{3}$ &148 &1969  \\ \hline
$F\bar{4}3m$ &216 & 898  &$I4/mmm$ & 139 &6162  \\ \hline
$P6_3/mmc$ &194 & 5599  &$Fd\bar{3}m$ & 227 &3292  \\ \hline
$P\bar{3}m1$ & 164 &1191  &$Pnma$  &62 &2527  \\ \hline
$P6/mmm$ &191& 2214  &$R\bar{3}m$ & 166 &1479  \\ \hline
$I4/mcm$ &140 & 433  &$P6_3mc$ &186 &692  \\ \hline
$R\bar{3}c$ &167& 1246  &$P4/nmm$ &129 &992  \\ \hline
\end{tabular}
\caption{20 space groups and their frequency in dataset \textbf{MIO}.}
\label{adx-tab:mio}
\end{table}
\endgroup

\begingroup
\setlength{\tabcolsep}{8pt} 
\renewcommand{\arraystretch}{1.2} 
\begin{table}[ht]
\centering
\begin{tabular}{|c|c|c|c|c|c|}
\hline
SG & SG Id &\# samples & SG & SG Id &\# samples \\ \hline
$P4/mmm$ &123 &317   &$Immm$ &71 &59  \\ \hline
$Fm\bar{3}m$ &225 & 675  &$Cmcm$ & 63 &507  \\ \hline
$I4_1/amd$ &141 &168  &$I\bar{4}2d$ &122  &482  \\ \hline
$P4/nmm$ &129 &719  &$R\bar{3}$ &148  &374  \\ \hline
$F\bar{4}3m$ &216 &60  &$I4/mmm$ &139 &768  \\ \hline
$P6_3/mmc$ &194 &1713  &$Fd\bar{3}m$ &227 &239  \\ \hline
$P\bar{3}m1$ &164 &674  &$Pnma$ &62  &1386  \\ \hline
$P6/mmm$ &191 &281  &$R\bar{3}m$ &166 &576  \\ \hline
$I4/mcm$ &140 &81  &$P6_3mc$ &186  &151  \\ \hline
$R\bar{3}c$ &167 &211  &$Pm\bar{3}m$ &221 &0  \\ \hline
\end{tabular}
\caption{20 space groups and their frequency in dataset \textbf{TST}.}
\label{adx-tab:tst}
\end{table}
\endgroup

We use first, second, and four criteria above to select materials in new released OQMD and the distribution of materials in 20 space groups is shown in Table~\ref{adx-tab:tst}. 9441 materials are chosen and space group $Pm\bar{3}m$ does not have new released materials.

\section{Model Details}\label{adx:C}


\subsection{Implementation Hyperparameters for training \textbf{PGCGM}}
Table~\ref{adx-tab:hyp} shows the hyper-parameters in \textbf{PGCGM}. We use $\mathit{\lambda}_\mathit{1}=1$ and $\mathit{\lambda}_\mathit{2}=1$ for atom distance losses. We use \textit{Property distribution} to select best atom dist bound $\mathit{\phi}$s combination and then using best $\mathit{\phi}$s, we optimize the best base coordinates and average full coordinates loss coefficients $\mathit{\lambda}_\mathit{1}$ and $\mathit{\lambda}_\mathit{2}$. Table~\ref{adx-tab:atom_dist} and Table~\ref{adx-tab:coord} show the performance with different settings. We use 9 different combinations of $\mathit{\phi}$s and the best average \textit{Property distribution} is achieved when $\mathit{\phi}$s are $(0.3, 7.5, 0.15, 7.5)$ as shown in~\ref{adx-tab:atom_dist}. With best $\mathit{\phi}$s, we add coordinates based losses for the generator and the best $\mathit{\lambda}_\mathit{1}$ and $\mathit{\lambda}_\mathit{2}$ are 0.001 and 0.01 averagely as shown in Table~\ref{adx-tab:coord}.

\begin{table}[ht]
\centering
\begin{tabular}{|cc|c|}
\hline
\multicolumn{2}{|c|}{Hyper-parameters} &Values  \\ \hline
\multicolumn{1}{|c|}{\multirow{3}{*}{Adam optimizer}} &learning rate  &0.0002  \\ \cline{2-3} 
\multicolumn{1}{|c|}{}                  &$\mathit{\beta}_\mathit{1}$  &0.5  \\ \cline{2-3} 
\multicolumn{1}{|c|}{}                  &$\mathit{\beta}_\mathit{2}$  &0.5  \\ \hline
\multicolumn{2}{|c|}{batch size} &8192  \\ \hline
\multicolumn{2}{|c|}{gradient penalty coefficient} & 10 \\ \hline
\multicolumn{2}{|c|}{\# of iterations of D per G iteration} &5  \\ \hline
\multicolumn{2}{|c|}{low bound for inter atom dist $\mathit{\phi}_\mathit{inter}^\mathit{lower}$} &0.3  \\ \hline
\multicolumn{2}{|c|}{upper bound for inter atom dist $\mathit{\phi}_\mathit{inter}^\mathit{upper}$} &7.5  \\ \hline
\multicolumn{2}{|c|}{low bound for intra atom dist $\mathit{\phi}_\mathit{intra}^\mathit{lower}$} &0.15 \\ \hline
\multicolumn{2}{|c|}{upper bound for intra atom dist $\mathit{\phi}_\mathit{intra}^\mathit{upper}$} &7.5  \\ \hline
\multicolumn{2}{|c|}{inter dist loss coefficient $\mathit{\lambda}_\mathit{1}$} &1.0  \\ \hline
\multicolumn{2}{|c|}{intra dist loss coefficient $\mathit{\lambda}_\mathit{2}$} &1.0  \\ \hline
\multicolumn{2}{|c|}{base coord diff loss coefficient $\mathit{\lambda}_\mathit{3}$} &0.001  \\ \hline
\multicolumn{2}{|c|}{avg. full coord loss coefficient $\mathit{\lambda}_\mathit{4}$} &0.1  \\ \hline
\end{tabular}
\caption{Hyper-parameters for training.}
\label{adx-tab:hyp}
\end{table}

\begin{table}[ht]
\centering
\begin{tabular}{|c|c|c|c|c|}
\hline
$\mathit{\phi}$s &minD  & maxD &density&avg.  \\ \hline
(0.3, 7.8, 0.0009, 30.5) &{0.220} &0.846  &1.481 & 0.849 \\ \hline
(0.3, 12.5, 0.15, 25.0) &0.256 &1.703  &1.770 &1.243 \\ \hline
(0.3, 15.0, 0.15, 25.0) &0.228 &1.879  &2.139 &1.415 \\ \hline
(0.3, 7.5, 0.15, 12.5) &0.401 &0.834  &{0.548} &0.594 \\ \hline
(0.3, 7.5, 0.15, 20.0) &0.301	&1.000  &1.176 &0.826 \\ \hline
\textbf{(0.3, 7.5, 0.15, 7.5)} &\textbf{0.354} &\textbf{0.512}  &\textbf{0.757} &\textbf{0.541} \\ \hline
(0.3, 2.75, 0.15, 2.75) &0.573	&2.157  &3.214 &1.981 \\ \hline
(0.3, 5.0, 0.15, 5.0) &0.424 &0.590  &0.721 &0.578 \\ \hline
(0.3, 2.0, 0.15, 2.0) &0.728	&2.322  &3.848 &2.299 \\ \hline
\end{tabular}
\caption{Choose the best $\mathit{\phi}$s $(\mathit{\phi}_\mathit{inter}^\mathit{lower},\mathit{\phi}_\mathit{inter}^\mathit{upper},\mathit{\phi}_\mathit{intra}^\mathit{lower},\mathit{\phi}_\mathit{intra}^\mathit{upper})$ when adding dist losses.}
\label{adx-tab:atom_dist}
\end{table}

\begin{table}[ht]
\centering
\begin{tabular}{|c|c|c|c|c|}
\hline
$(\mathit{\lambda}_\mathit{1},\mathit{\lambda}_\mathit{2})$ &minD  & maxD &density&avg.  \\ \hline
(0.001, 0.0001) &0.301	&0.594  &0.993&0.629  \\ \hline
(0.001, 0.001) &0.258 &1.103  &0.823&0.728  \\ \hline
(0.0001, 0.0001) &0.299	&1.346  &1.206&0.950  \\ \hline
(0.0001, 0.001) &0.367 &0.770  &0.728&0.622  \\ \hline
(0.01, 0.001) &0.337 &1.032  &1.440&0.936  \\ \hline
(0.01, 0.01) &0.203 &1.147  &2.17&1.173  \\ \hline
\textbf{(0.001, 0.01)} &\textbf{0.308}	&\textbf{0.504}  &\textbf{0.689}&\textbf{0.500}  \\ \hline 
(0.01, 0.0001) &0.251	&0.991  &0.942&0.728  \\ \hline
(0.1, 0.1) &{0.159}	&1.359  &2.918&1.479 \\ \hline
\end{tabular}
\caption{Choose the best $\mathit{\lambda}_\mathit{1}$ and $\mathit{\lambda}_\mathit{2}$ when adding coordinates based losses.}
\label{adx-tab:coord}
\end{table}

\subsection{Model Structures}
Table~\ref{adx-tab:d_model} and~\ref{adx-tab:g_model} show the detailed architectures of discriminator and generator.

\begin{table}[ht]
\centering
\begin{tabular}{|cc|}
\hline
\multicolumn{2}{|c|}{Discriminator Configuration}                     \\ \hline
\multicolumn{1}{|c|}{\textbf{Mat-$3\times 8\times 8$}} & \multirow{3}{*}{} \\ \cline{1-1}
\multicolumn{1}{|c|}{C2D-16-2} &                   \\ \cline{1-1}
\multicolumn{1}{|c|}{C2D-32-2} &                   \\ \cline{1-1}
\multicolumn{1}{|c|}{C2D-64-2} &                   \\ \hline
\multicolumn{1}{|c|}{C2D-96-2} & $\mathbf{SymOp}$-$192\times 4\times 4$ \\ \hline
\multicolumn{1}{|c|}{C2D-128-2} & C2D-64-2\\ \hline
\multicolumn{1}{|c|}{C2D-192-2} & C2D-128-2 \\ \hline
\multicolumn{1}{|c|}{C2D-256-2} & C2D-256-2\\ \hline
\multicolumn{2}{|c|}{CAT-512}                     \\ \hline
\multicolumn{2}{|c|}{FC-265}                     \\ \hline
\multicolumn{2}{|c|}{FC-1}                     \\ \hline
\end{tabular}
\caption{Discriminator configuration. \textbf{Mat} is the input material representations with shape of $3\times 8\times 8$. $\mathbf{SymOp}$ is the zero-padded symmetric operation matrix for space groups of materials. The 2D convolutional layer parameters are denoted as "C2D-$<$number of channels$>$-$<$receptive field size$>$". The fully connected layer parameters are denotes as "FC-$<$number of neurons$>$".  The concatenation is denoted as "CAT-$<$number of neurons$>$". We use \textit{LeakyReLU} as the activation function after each layer except for the last layer. The negative slope for it is 0.2.}
\label{adx-tab:d_model}
\end{table}

\begin{table}[ht]
\centering
\begin{tabular}{|cccc|}
\hline
\multicolumn{4}{|c|}{Generator Configuration}  \\ \hline
\multicolumn{1}{|c|}{} & \multicolumn{2}{c|}{\multirow{3}{*}{}}  & $\mathbf{ElemProp}$-$23\times 3$ \\ \cline{1-1} \cline{4-4} 
\multicolumn{1}{|c|}{$\mathbf{SymOp}$-$192\times 4\times 4$} & \multicolumn{2}{c|}{}&C1D-64-2  \\ \cline{1-1} \cline{4-4} 
\multicolumn{1}{|c|}{C2D-64-2} & \multicolumn{2}{c|}{}                   &C1D-128-2  \\ \hline
\multicolumn{1}{|c|}{C2D-128-2} & \multicolumn{2}{c|}{\textbf{Z}-128} &flatten  \\ \hline
\multicolumn{1}{|c|}{C2D-256-2} & \multicolumn{2}{c|}{FC-256}&FC-256  \\ \hline
\multicolumn{2}{|c|}{~~~~~~~~~~~~~~CAT-512~~~~~~~~~~~~~~~}& \multicolumn{2}{c|}{CAT-512} \\ \hline
\multicolumn{2}{|c|}{FC-128}& \multicolumn{2}{c|}{TC2D-1024-2} \\ \hline
\multicolumn{2}{|c|}{FC-64}& \multicolumn{2}{c|}{TC2D-512-2} \\ \hline
\multicolumn{2}{|c|}{FC-32}& \multicolumn{2}{c|}{TC2D-256-1} \\ \hline
\multicolumn{2}{|c|}{FC-16}& \multicolumn{2}{c|}{TC2D-128-1} \\ \hline
\multicolumn{2}{|c|}{FC-3}& \multicolumn{2}{c|}{TC2D-64-1} \\ \hline
\multicolumn{2}{|c|}{\multirow{2}{*}{output: $\mathbf{P}^{*}$-3 }}     & \multicolumn{2}{c|}{TC2D-3-1} \\ \cline{3-4} 
\multicolumn{2}{|c|}{}                      & \multicolumn{2}{c|}{output: $\mathbf{B}$ - $3\times 3\times 3$} \\ \hline
\end{tabular}
\caption{Generator configuration. $\mathbf{SymOp}$ is the zero-padded symmetric operation matrix for space groups of materials. $\mathbf{Z}$ is the random noise with shape of 128 and it shared by two branches for generating unit cell length $\mathbf{P}^{*}$ and three set of base atom sites $(\mathbf{B}_\mathit{fake}^\mathit{0},\mathbf{B}_\mathit{fake}^\mathit{1},\mathbf{B}_\mathit{fake}^\mathit{2})$. The 2D convolutional layer parameters are denoted as "C2D-$<$number of channels$>$-$<$receptive field size$>$". The 2D deconvolutional layer parameters are denoted as "TC2D-$<$number of channels$>$-$<$receptive field size$>$". The fully connected layer parameters are denotes as "FC-$<$number of neurons$>$".  The concatenation is denoted as "CAT-$<$number of neurons$>$". We use batch normalization and \textit{ReLU} after each layer except for the last layers of two branches. They are followed by a \textit{Tanh} activation to generate lengths and atom coordinates.}
\label{adx-tab:g_model}
\end{table}

\section{DFT configuration}\label{adx:D}

The structures were optimized by density functional theory (DFT)
that were carried out with Vienna ab initio simulation package (VASP).
The structure optimization convergence criteria of force and
energy are ${10}^{-4}$ eV/A and ${10}^{-7}$ eV, respectively. VASP runs were performed with full degree of freedom in terms of allowing the atomic coordinates, lattice size, lattice constant, and lattice shape to change to reach the convergence criteria of force and energy in the structure optimization process. The Perdew–Burke–Ernzerhof (PBE) of the generalized gradient approximation (GGA) was used for exchange–correlation functional. The kinetic energy cutoff was set to be 520 eV for the electronic wavefunction having a plane wave basis set which was obtained using the projector augmented-wave method. The Monkhorst–pack k-mesh grids selected to sample the Brillouin zone in the calculations were determined depending on the lattice constants. The product of the number of k-meshes in one direction and the lattice constant (measured in Angstrom) in the same direction is roughly set as 60, which is dense enough for structure optimization.

\bibliography{supp-ref}